\documentclass[a4paper,11pt]{article}
\pdfoutput=1 % if your are submitting a pdflatex (i.e. if you have
\usepackage{jcappub} % for details on the use of the package, please
\usepackage[T1]{fontenc} % if needed

\usepackage{braket}
\usepackage{adjustbox}
\usepackage{graphicx,epsfig}
\usepackage{amsmath}
\usepackage {amssymb}
\usepackage {longtable}
\usepackage{multirow}
 \usepackage{mathrsfs}

\usepackage{dcolumn}
\usepackage{bm}
\usepackage{amsfonts}
\usepackage{subfigure}
\usepackage{color}
\usepackage{relsize}

\newcommand{\be}{\begin{equation}}
\newcommand{\ee}{\end{equation}}
\newcommand{\bea}{\begin{eqnarray}}
\newcommand{\eea}{\end{eqnarray}}

\title{\boldmath Noether Symmetry in a Non-Metricity Theory with a Boundary Term: Exact Solutions and Bayesian Cosmological Constraints}

\author[a]{Pooja Vishwakarma,}
\author[b]{L K Duchaniya,\note{Corresponding author.}}
\author[b]{Yu Shi}

\affiliation[a]{Mukesh Patel School of Technology Management \& Engineering, SVKM's NMIMS, Mumbai, India.}
\affiliation[b]{Wilczek Quantum Center, Shanghai Institute for Advanced Studies, Shanghai 201315  \\University of Science and Technology of China, Hefei 230026, China.}

\emailAdd{poojavishwakarma041993@gmail.com}
\emailAdd{duchaniya98@gmail.com}
\emailAdd{yu\_shi@ustc.edu.cn}

\abstract{ We investigate the cosmological implications of a power-law $f(Q, B)$ gravity model, where $Q$ denotes the non-metricity scalar and $B$ is the associated boundary term. We employ the Noether symmetry approach to identify the admissible functional form of the gravitational Lagrangian and to obtain the corresponding conserved quantities and exact cosmological solutions. Using the exact solution, we find that for this model parameter condition $\alpha+\beta>\frac{3}{2}$, the model exhibits the accelerating phase of the Universe. Furthermore, we have also discussed the model with recent observational data using a Bayesian Markov Chain Monte Carlo analysis. We consider different combinations of Cosmic Chronometers (CC), Pantheon+ \& SH0ES, and DESI DR2 observations to constrain the model parameters and the Hubble constant $H_0$. The inferred value of $H_0$ depends on the adopted dataset combination, ranging from a higher value for the CC+Pantheon+ combination to a lower value for CC+DESI DR2, with the latter yielding $H_0=67.5^{+1.4}_{-1.6}\,\mathrm{km\,s^{-1}\, Mpc^{-1}}$, consistent with the Planck 2018 CMB-preferred value within the standard $\Lambda$CDM framework. The inferred \(H_0\) is strongly dataset dependent, spanning the range between the early-Universe and locally calibrated determinations. We further examine the posterior covariance and correlation matrices to identify parameter degeneracies and assess the robustness of the MCMC constraints. The background evolution also exhibits late-time accelerated expansion and remains compatible with the observational constraints. These results demonstrate that the  $f(Q, B)$ model provides a viable alternative description of late-time cosmology.}

\begin{document}

\maketitle
\flushbottom

\section{Introduction} \label{SEC_Introduction}
The $\Lambda$CDM cosmological model has been extremely successful in explaining a wide variety of observations, from the anisotropies observed in the cosmic microwave background (CMB) to the large-scale distribution of matter and the accelerated expansion of the Universe in its later stages \cite{Peebles:2002gy, Copeland:2006wr, Riess:1998cb, Perlmutter:1998np}. Even with this phenomenological success, a number of observational and theoretical discrepancies have arisen in recent years, indicating that the standard cosmological framework might be lacking \cite{Weinberg:1988cp, DiValentino:2020zio}. The most critical of these inconsistencies is the Hubble tension, which refers to the ongoing conflict between the Hubble constant value determined from early-Universe observations and the value obtained from late-time measurements \cite{Planck:2018vyg,Riess_2021L6,Riess_2022panplus}. Specifically, the SH0ES distance-ladder measurement gives $H_0 = 73.04 \pm 1.04\, \text{km} \, \text{s}^{-1} \text{Mpc}^{-1}$ \cite{Riess_2022panplus}, while the 2018 Planck analysis based on $\Lambda$CDM provides $H_0 = 67.4 \pm 0.5\, \text{km} \, \text{s}^{-1} \text{Mpc}^{-1}$ \cite{Planck:2018vyg}. The statistical significance of this difference has now surpassed the threshold where it can be attributed solely to statistical fluctuations, prompting an investigation of new physics beyond the conventional cosmological model \cite{Abdalla:2022yfr,Verde_2019tensionearlylate}. These discrepancies imply that the $\Lambda$CDM model may not completely encompass all the missing aspects of physics. The finding of the accelerating expansion of the Universe is one of the most astonishing discoveries in cosmology, suggesting either a failure of Einstein's gravitational theory on a cosmological scale or the prevalence of dark energy with unusual characteristics. The acceleration of the Universe has sparked efforts to uncover its source, with experiments measuring expansion and growth with precision of 1\% or better. General relativity (GR) is nonrenormalizable on microscopic scales \cite{hooft1974, Goroff1985}, and the lack of a consistent theory of quantum gravity requires modifications or extensions to GR. The conventional method formulates gravity theories that include GR in particular situations while adding extra degrees of freedom \cite{Capozziello:2011et,Nojiri_2007_04, Ferraro:2008ey,Bahamonde:2017ize, Jimenez_2018_jcap}. These theories stem from basic considerations of gravitational field configurations and principles \cite{B_hmer_2021, CANTATA:2021ktz}, emphasizing the importance of preserving the principle of equivalence at quantum levels and tackling causal and geodesic structures. The necessity of the metric remains crucial when treating gravity as a gauge theory.

Recently studies in cosmology have renewed curiosity about the potential for dynamic dark energy. Specifically, the most recent findings from the Dark Energy Spectroscopic Instrument (DESI) have contributed to this interest. The CMB data represent the condition of the Universe at redshift z$\approx$1090, just 400,000 years after the Big Bang. Baryon acoustic oscillations (BAO) obtained from galaxy surveys restrict the Universe's expansion history within the range of $0.1<z<4.2$, which includes both the matter-dominated era and the recent phase of cosmic acceleration. Observations of Type Ia supernovae (SNe Ia) were employed to reveal dark energy and enhance the understanding of the expansion history at lower redshifts. The combined findings from these cosmological observations across the evolving Universe offer quantitative insights into the temporal changes in dark energy density. Recently, the DESI collaboration introduced a cosmological analysis based on the latest BAO measurements from its Data Release 2 (DR2) \cite{Abdul_Karim_2025_Desi}. This analysis suggests that the inconsistencies among datasets are becoming more pronounced within the $\Lambda$CDM framework, proposing the concept of evolving dark energy as a possible solution and also providing updated boundaries on the neutrino mass \cite{elbers2025constraintsneutrinophysicsdesi}. In particular, while the DESI DR2 BAO findings generally agree with the $\Lambda$CDM cosmological model, they expose a discrepancy of 2.3$\sigma$ \cite{Abdul_Karim_2025_Desi} when compared to Planck CMB data. On the other hand, a time-varying dark energy component is supported by the combined evaluation of these consistent datasets. This trend can be inferred through straightforward CPL parametrizations \cite{Linder_2003_para} or various methods for reconstructing dark energy \cite{lodha2025extendeddarkenergyanalysis}. Utilizing the $\omega_0 \omega_a$CDM framework, the combination of DESI BAO DR2 with fluctuations in CMB temperature and polarization, in addition to CMB lensing, indicates that evolving dark energy is preferred at a significance level of $3.1\sigma$. This preference fluctuates between $2.8\sigma$ and $4.2\sigma$ when SNe Ia data are incorporated, depending on the specific SNe sample used \cite{Abdul_Karim_2025_Desi} (this preference is also supported by the Dark Energy Survey's integrated analysis of BAO and SNe \cite{descollaboration2026darkenergysurveyimplications}).

A large variety of approaches has been proposed to explain the late-time accelerated expansion of the Universe and cosmological tensions, including early dark energy models, interacting dark energy scenarios and modifications of GR. New categories of modified gravity theories have come to light, integrating curvature, torsion, and non-metricity scalars. These categories emerge even though the unaltered theories are mathematically equivalent at the level of equations. The crucial factor is the distinction between the torsion scalar $T$ and the non-metricity scalar $Q$, which diverges from the standard Levi–Civita Ricci scalar $R$ of GR due to additional components: $R=-T+B$ and $R=Q+B$, respectively, where $B$ represents the boundary term. A geometric trinity representing second-order gravity can be seen in $R$, $B- T$, and $Q + B$, while $f(R)$, $f(B-T)$, and $f(Q + B)$ can be considered a geometric trinity of fourth-order gravity \cite{jimenez2019geometricaltrinitygravity, Capozziello_2022ab}. The modified gravity theories such as \cite{Ferraro:2008ey,Bengochea:2008gz,Linder:2010py,Basilakos:2013rua,Finch:2018gkh,Bahamonde:2017ize,Basilakos:2018arq,Gonzalez-Espinoza:2020jss, Gonzalez_Espinoza_2020A, Duchaniya_2023, Duchaniya:2022rqu,duchaniya2025latetimephenomenaftmathcalt, Vishwakarma_2024, Lohakare_2024, Alwan_2025, Lohakare_2026, Lohakare:2026noa, pati2026cosmologicalimplicationsftb}, scalar-torsion theories \cite{Hohmann:2018rwf, Sakstein:2017xjx, Bartolo_1999, Horndeski:2024sjk, Kadam:2022lgq, Duchaniya_2024cqg, Gonzalez-Espinoza:2021mwr, Fomin_2025pdu}, early dark energy scenarios \cite{Poulin:2018cxd, Sakstein:2019fmf, Gogoi:2020qif, Niedermann2020, Murgia2021, Freese:2021rjq, Herold:2022iib} and interacting dark sectors \cite{DiValentino:2017iww, An:2017crg, Yang:2018qmz, Yang:2018uae,  Gao:2021xnk, duchaniya2025latetimeaccelerationstructureformation}. However, many proposed models introduce substantial modifications to the standard cosmological framework or require additional degrees of freedom and fine-tuning of model parameters, which may reduce their predictive power. This has motivated increasing interest in mechanisms that primarily operate in the late Universe and can modify the observed cosmic expansion history while remaining consistent with the established cosmological framework.

In the present work, we investigate the Noether symmetry approach within the framework of f(Q, B) gravity and perform a comprehensive observational analysis of the resulting cosmological model using various combinations of cosmological datasets. Using Noether symmetry, we will examine possible scenarios for cosmological evolution arising from specific models within this category of theories. Noether symmetries are a valuable resource for addressing dynamical equations in cosmology; beyond that, they enable the development of models inspired by fundamental principles. This enhances the rationale for investigating intricate systems of equations of motion. By the Noether symmetry associated with a specific minisuperspace cosmological model, we indicate that there is a vector field $\mathcal{X}$ that serves as the infinitesimal generator of the symmetry on the tangent space of the configuration space, such that the Lie derivative of the Lagrangian concerning this vector field equals zero. The Noether symmetry approach detailed in Refs .~\cite {PhysRevD.42.1091, PhysRevD.46.1391, Capozziello_2000, Dialektopoulos_2022, Bahamonde_2019, Kucukakca_2013,Bahamonde_2017n, Kadam_2023ns, Duchaniya_2023noet}. Observational analyses of$f(Q, B)$ cosmology have been investigated in previous studies, including the dynamical-system and Bayesian analyses of \cite{Lohakare_2024}, as well as the related $f(T, B)$ framework studied in \cite{pati2026cosmologicalimplicationsftb}. In the present work, we consider a power-law form of the $f(Q, B)$ gravitational Lagrangian and investigate its cosmological dynamics through the Noether symmetry approach. The existence of the Noether symmetry provides a systematic way to identify conserved quantities and reduce the cosmological equations, allowing us to obtain an exact analytical solution for the background evolution. We subsequently confront this theoretically motivated $f(Q, B)$ model with observational data to examine its cosmological viability, incorporating the latest DESI DR2 BAO measurements together with Type Ia supernovae and cosmic chronometers data set.

We organize the work as follows: in Sec .~\ref {SECII_Symmmathmatical}, we briefly discuss symmetric teleparallel and $f(Q, B)$ gravity mathematical formalism, together with the formulation of the Friedmann equations for this setting. In Sec .~\ref {SEC:Lagrangian formalism}, we obtain the point-like Lagrangian and derive the Noether equations using the Euler-Lagrange equations in configuration space $\mathcal{Q}=(a, Q, B)$, leading to the cosmological equations of motion (\ref{Eq:rhode}--\ref{Eq:Pde}). In Sec .~\ref {SEC:Noether Symmetry}, we introduce the concept of Noether symmetries, and we have also obtained the exact solution of the Friedmann equation by using a particular form of $f(Q, B)$. In Sec .~\ref{SEC:cosmologicalobsevation}, we present Bayesian analysis using different data sets and study the late-time cosmology for each data-set combination. Finally, we conclude with a summary of the main results in Sec .~\ref {Sec:conclusion}.

\section{ Symmetric teleparallel gravity}\label{SECII_Symmmathmatical}

We explore a gravitational framework defined by a four-dimensional metric tensor $g_{\mu\nu}$ and the covariant derivative $\nabla_{\mu}~$, formulated through the generic connection $\Gamma_{\mu\nu}^{\zeta}\,$. Within the framework of Symmetric Teleparallel General Relativity (STGR), the connection $\Gamma_{\mu\nu}^{\zeta}\,$ is flat and torsion-free. As a result, we determine that $R_{\;\eta\mu\nu}^{\zeta}=0$ and $\mathrm{T}_{\mu\nu}^{\eta}=0$. Additionally, it preserves the symmetries inherent in the metric tensor $g_{\mu\nu}$. Autoparallels are defined as explained by Obukhov  and Puetzfeld \cite{Obukhov_2021}. 
\begin{equation}
\frac{d^{2}x^{\mu}}{ds^{2}}+\Gamma_{\zeta\nu}^{\mu}\frac{dx^{\zeta}}{ds}\frac{dx^{\nu}}{ds}=0. \label{eq: 1}
\end{equation}

The characteristics of the geometry are dictated by the connection, denoted as $\Gamma_{\mu\nu}^{\zeta}$. The Riemann tensor can be expressed in terms of the general connection.
\begin{equation}
R_{\;\eta\mu\nu}^{\zeta} = \frac{\partial\Gamma_{\;\eta\nu}^{\zeta}% 
}{\partial x^{\mu}}-\frac{\partial\Gamma_{\;\eta\mu}^{\zeta}}{\partial x^{\nu}}+\Gamma_{\;\eta\nu}^{\sigma}\Gamma_{\;\mu\sigma}^{\zeta}%
        -\Gamma_{\;\eta\mu}^{\sigma}\Gamma_{\;\mu\sigma}^{\zeta},
        \label{eq: 2}
\end{equation}
the torsion tensor
\begin{equation}
\mathrm{T}_{\mu\nu}^{\eta}=\Gamma_{\;\mu\nu}^{\eta}-\Gamma_{\;\nu\mu}^{\eta}, \label{eq: 3}
\end{equation}
and the nonmetricity tensor 
\begin{eqnarray}
Q_{\eta\mu\nu}=\nabla_{\eta}g_{\mu\nu}=\frac{\partial g_{\mu\nu}%
        }{\partial x^{\eta}}-\Gamma_{\;\eta\mu}^{\sigma}g_{\sigma\nu}%
        -\Gamma_{\;\eta\nu}^{\sigma}g_{\mu\sigma}. \label{eq: 4}
\end{eqnarray}

In symmetric teleparallel theory, it is always possible to select an appropriate diffeomorphism that eliminates the general affine connection $\Gamma_{\mu\nu}^{\zeta}$, referred to as the coincident gauge \cite{Jim_nez_2018coin}. As a result, the covariant derivative simplifies to the partial derivative, and the principles of symmetric teleparallelism (i.e., zero curvature and torsion) dictate that the general connection turns into the Levi-Civita connection, which inherently possesses symmetry.

In the Teleparallel Equivalent of General Relativity (TEGR), the conventional connection $\Gamma_{\mu\nu}^{\zeta}$ is substituted by the antisymmetric Weitzenböck connection. This leads to the result that $R_{;\eta\mu\nu}^{\zeta} = 0$ and $Q_{\eta\mu\nu}=0$. Within this framework, the torsion scalar $T$ is established as the key geometric entity in teleparallel gravity. 

The nonmetricity scalar $Q$ can be defined as \cite{nester1999symmetricteleparallelgeneralrelativity}
\begin{equation}
Q=Q_{\eta\mu\nu}P^{\eta\mu\nu} \, , \label{eq: 5}
\end{equation}
the nonmetricity conjugate  $P^{\eta\mu\nu}$  can be defined as
\begin{equation}
P_{\;\mu\nu}^{\eta}=-\frac{1}{4}Q_{\;\mu\nu}^{\eta}+\frac{1}{2}        Q_{(\mu\phantom{\eta}\nu)}^{\phantom{(\mu}\eta\phantom{\nu)}}+\frac{1}{4}\left(  Q^{\eta}-\bar{Q}^{\eta}\right)  g_{\mu\nu}-\frac{1}
        {4}\delta_{\;(\mu}^{\eta}Q_{\nu)}, \label{eq: 6}
\end{equation}

the expressions $Q_{\mu}=Q_{\mu\nu}^{\phantom{\mu\nu}\nu}$ and $\bar{Q}_{\mu}=Q_{\phantom{\nu}\mu\nu}^{\nu\phantom{\mu}\phantom{\mu}}$ are utilized in this context. The Ricci scalar $R$ is associated with the Levi-Civita connection $\tilde{\Gamma}_{\mu\nu}^{\zeta}$ of the metric tensor $g_{\mu\nu}$. The nonmetricity scalar $Q$, in the case of a symmetric and flat connection $\Gamma_{\mu\nu}^{\zeta}$, differs from $R$ by a boundary term $B$, which is expressed as $B=R-Q$. The gravitational action for STGR is expressed as follows,
\begin{equation}
\int d^{4}x\sqrt{-g}Q\simeq\int d^{4}x\sqrt{-g}(R - B )\, , \label{eq: 7}
    \end{equation}

The boundary term ($B$) does not affect the field equations under appropriate boundary conditions, which means that STGR is dynamically equivalent to General Relativity, yielding the same cosmological equations. However, this equivalence breaks down when nonlinear aspects of the nonmetricity scalar $Q$ are added, as seen in $f(Q)$ gravity within the gravitational action. The action formula for symmetric teleparallel $f(Q)$ gravity can be  expressed as
\begin{equation}
S_{f\left(  Q\right)  }=\int d^{4}x\sqrt{-g}f(Q). \label{eq: actionfQ}
\end{equation}

%%%%%%%%%%%%%%%%%%%%%%%%%%%%%%%%%%%%%%%%%%%%%%%%%%
    %%%%%%%%%%%%%%%%%%%%%%%%%%%%%%%%%%%%%%%%%%%%%%%%%%

\subsection{\texorpdfstring{$f(Q, B)$}{} cosmology} \label{Sec_fQBmath}

An extension of the $f(Q)$ gravity theory is the $f(Q, B)$ theory \cite{Capozziello_2022ab, De_2024,Lohakare_2024}, which incorporates a boundary term into the gravitational action. The action for $f(Q, B)$ gravity is expressed as
\begin{equation} \label{eq: action_fQB}
S = \int  d^{4}x \,\sqrt{-g} \left[\frac{1}{2 \kappa} f(Q, B)+ \mathcal{L}_{m} +\mathcal{L}_{r}\right]  , 
\end{equation}

where $g$ denotes the determinant of the metric tensor $g_{\mu\nu}$ and $\kappa = 8\pi G $ is the gravitational coupling constant. The quantity $(\mathcal{L}_{m})$ represents the Lagrangian density of the matter fields, while $(\mathcal{L}_{r})$ denotes the Lagrangian density of the radiation component. The integration is performed over the four-dimensional spacetime manifold. Varying the action \eqref{eq: action_fQB} with respect to metric tensor $g_{\mu\nu}$ leads the following field equation \cite{De_2024}
\begin{eqnarray}
\kappa T_{\mu\nu}&=&-\frac f2g_{\mu\nu} +\frac2{\sqrt{-g}}\partial_\eta \left(\sqrt{-g}f_Q P^\eta{}_{\mu\nu}\right)+(P_{\mu\alpha\beta}Q_\nu{}^{\alpha\beta}-2P_{\alpha\beta\nu}Q^{\alpha\beta}{}_\mu)f_Q \nonumber\\&& +\left(\frac B2 g_{\mu\nu}-\nabla_{\mu}\nabla_{\nu} +g_{\mu\nu}\nabla^\alpha\nabla_\alpha-2P^\eta{}_{\mu\nu}\partial_\eta \right)f_B\,,\label{eqn:FE1-pre}
\end{eqnarray}
this can be defined in a covariant way
\begin{equation}
        \kappa T_{\mu\nu}=-\frac f2g_{\mu\nu}+2P^\eta{}_{\mu\nu}\nabla_\eta(f_Q-f_B)
        +\left(G_{\mu\nu}+\frac Q2g_{\mu\nu}\right)f_Q
        +\left(\frac B2g_{\mu\nu}-\nabla_{\mu}\nabla_{\nu}
        +g_{\mu\nu}\nabla^\alpha\nabla_\alpha \right)f_B\,. \label{eq:11}
\end{equation}
An effective stress-energy tensor can be defined in the following way
\begin{equation} \label{T^eff}
        T^{\text{eff}}_{\mu\nu} =  T_{\mu\nu}+ \frac 1{\kappa}\left[\frac f2g_{\mu\nu}-2P^\eta{}_{\mu\nu}\nabla_\eta(f_Q-f_B)
        -\frac {Qf_Q}2g_{\mu\nu}\right.
        \left.-\left(\frac B2g_{\mu\nu}-\nabla_{\mu}\nabla_{\nu}
        +g_{\mu\nu}\nabla^\alpha\nabla_\alpha \right)f_B\right]\,.
\end{equation}
To create an equation that resembles that of GR
\begin{align}
G_{\mu\nu}=\frac{\kappa}{f_Q}T^{\text{eff}}_{\mu\nu}\,. \label{eq: 13}
\end{align}

We now examine the cosmological formulation of the $f(Q,B)$ gravity theory. Throughout this work, the background geometry is assumed to be a homogeneous, isotropic and spatially flat Friedmann-Lema\^{i}tre-Robertson-Walker (FLRW) spacetime, described by the following line element in cartesian coordinates
\begin{equation}
ds^{2}=-dt^{2}+a^{2}(t)[dx^{2} + dy^{2}+dz^{2}],  \label{eq:FLRW metric}
\end{equation}
where $a(t)$ represents the scale factor, which measures the expansion rate of the Universe.

After this section, it has been shown that in the framework of $f(Q, B)$ gravity, one can derive an extra effective sector originating from geometry, as illustrated in Eq. \eqref{T^eff}. As a result, in a cosmological context, this term can be understood as an effective sector of dark energy represented by an energy-momentum tensor.
\begin{equation}
            T^{\text{DE}}_{\mu\nu}= \frac 1{f_Q}\left[\frac 
            f2g_{\mu\nu}-2P^\eta{}_{\mu\nu}\nabla_\eta(f_Q-f_B)
            -\frac {Qf_Q}2g_{\mu\nu}\right.
            \left.  -\left(\frac B2g_{\mu\nu}-\nabla_{\mu}\nabla_{\nu} + g_{\mu\nu} \nabla^\alpha \nabla_\alpha \right)f_B\right]\, , \label{eq: 15}
\end{equation}

\begin{align}
        R = & 6(2H^2 + \dot{H}), \quad Q = -6H^2,
        \quad B = 3(6H^2 +2\dot{H}). \label{eq: QRB}
\end{align}

In this scenario, we assume an affine connection that vanishes $(\Gamma_{\mu \nu}^{\eta}=0)$ while applying the coincident gauge. From this information, we can formulate our Friedmann equations as indicated below
\begin{eqnarray}
  3 H^2 &=&\kappa \left(\rho_{\text{m}}+\rho_{\text{r}} + \rho_{\text{DE}}\right)\, ,  \label{first_field_equation}\\
 -2 \dot{H}-3 H^2& =&\kappa \left(\frac{\rho_{\text{r}}}{3} + p_{\text{DE}}\right) \, . \label{second_field_equation}
\end{eqnarray}

In this context, $\rho_{\mathrm{m}}$, $\rho_{\mathrm{r}}$ and $\rho_{\mathrm{DE}}$, denote the energy densities associated with matter, radiation and dark energy, respectively and $p_{\mathrm{DE}}$ denote the pressure of the dark energy component, with dark energy being represented as a perfect fluid. The Hubble parameter is indicated by $H(t)=\frac{\dot{a}(t)}{a(t)}$. When assuming that non-relativistic matter and radiation evolve separately without interacting, their energy densities adhere to the standard conservation equations, $\dot{\rho}_{\mathrm{m}} + 3H\rho_{\mathrm{m}} = 0 $ and $\dot{\rho}_{\mathrm{r}} + 4H\rho_{\mathrm{r}} = 0,$ which apply to matter and radiation, respectively. Additionally, the effective density and pressure of dark energy are defined as follows   
\begin{eqnarray}
 \kappa \,\rho_{\text{DE}}&=&\left[6 H^2 f_Q-\frac{f}{2}+\left(9 H^2+3 \dot{H}\right) f_B-3 H \dot{f_B}\right]\,, \label{Eq:rhode}\\
  \kappa \,p_{\text{DE}}&=&\Big[2 \dot{H} f_Q-3 H^2\left(1-2 f_Q\right)+\frac{f}{2}+2 H \dot{f}_Q -\left(9 H^2+3 \dot{H}\right) f_B+\ddot{f_B}\Big].\label{Eq:Pde}
\end{eqnarray}

The evolution of the total equation-of-state (EoS) parameter, $\omega$, and the deceleration parameter $q$, plays a crucial role in characterizing the late-time dynamics of the Universe. These parameters are defined as follows
\begin{eqnarray} \label{omegatot} 
\omega&=& \frac{p_{m}+p_r+p_{DE}}{\rho_{m}+\rho_r+\rho_{DE}}\equiv -1-\frac{2\dot{H}}{3 H^{2}},\\
q &=& \frac{1}{2}(1+3 \omega)\,.\label{dece}
\end{eqnarray}

From cosmological observations, it is well established that the sign of the deceleration parameter $q$ determines the expansion behavior of the Universe. A positive value of $q$ corresponds to a decelerating expansion phase, whereas a negative value of $q$ indicates an accelerated expansion phase. Furthermore, the EoS parameter $\omega$ provides important information about the different evolutionary phases of the Universe, with various cosmic epochs being characterized by different ranges of $\omega$. 
%%%%%%%%%%%%%%%%%%%%%%%%%%%%%%%%%%%%%%%%%%%%%%%%%%%%%%%%%%%%%%%%%%%%%%%%%%%%%%%%%%%%%%%%%%%%%%%%%%%%%%%%%%%%%%%%%%%%%%%

\section{Lagrangian formalism of $f(Q,B)$ theory}   \label{SEC:Lagrangian formalism}

The point-like Lagrangian plays a vital role in the study of Noether symmetry. This section examines the Lagrangian formalism of the $f(Q, B)$ theory. By consulting sources like \cite{Capozziello_2000, Bahamonde_2019}, it is possible to formulate a canonical Lagrangian $\mathcal{L}=\mathcal{L}(a,\dot{a}, Q,\dot{Q}, B,\dot{B})$ to derive the cosmological equations within the FLRW metric. The configuration space is denoted by the set $\mathcal{Q}=\left\lbrace a,Q,B \right\rbrace $, while the associated tangent bundle, on which $\mathcal{L}$ is defined, is $\mathcal{T} \mathcal{Q} = \left\lbrace a,\dot{a},Q,\dot{Q},B,\dot{B} \right\rbrace $. The functions $a(t)$, $Q(t)$, and $B(t)$ are considered independent dynamical variables.
The method of Lagrange multipliers can be utilized to impose the constraint \(Q + 6\frac{\dot{a}^{2}}{a^{2}} = 0\) and \(B - 6\left(\frac{\ddot{a}}{a} + 2\left(\frac{\dot{a}}{a}\right)^{2}\right) = 0\) in the dynamics. By applying integration by parts, the Lagrangian $\mathcal{L}$ becomes analogous to the one presented in Ref. \cite{Capozziello_2000}, leading to the expression that follows:
\begin{equation}\label{eq:action_lagrangian1}
S=2\pi^{2}\int  a^{3}\left[f(Q,B)-\lambda_{1}\left( Q+6\frac{\dot{a}^{2}}{a^{2}}\right)-\lambda_{2}\left( B-6\left( \frac{\ddot{a}}{a}+2\left( \frac{\dot{a}}{a}\right)^{2} \right) \right)-\frac{\rho_{m0}}{a^{3}} \right] dt.
\end{equation}
Here, $ \lambda_{1} $ and $ \lambda_{2} $ are Lagrange multiplier.
The variation of this action with respect to $ Q $ and $ B $ gives
\begin{eqnarray}
(a^{3}f_{Q}-\lambda_{1})\delta Q=0 \rightarrow \lambda_{1}=a^{3}f_{Q}\\
(a^{3}f_{B}-\lambda_{2})\delta B=0 \rightarrow \lambda_{2}=a^{3}f_{B}
\end{eqnarray}
Thus, the action \eqref{eq:action_lagrangian1} can be rewritten as
\begin{equation}\label{eq:action_lagrangian2}
S=2\pi^{2}\int \left[ a^{3}f(Q,B)-a^{3}f_{Q}\left( Q+6\frac{\dot{a}^{2}}{a^{2}}\right)-a^{3}f_{B}\left( B-6\left( \frac{\ddot{a}}{a}+2\left( \frac{\dot{a}}{a}\right)^{2} \right) \right)-\rho_{m0}\right] dt.
\end{equation}
The point-like Lagrangian can be obtained as
\begin{equation}\label{Eq:main_Lagrangian}
\mathcal{L}(a,\dot{a},Q,\dot{Q},B,\dot{B})=a^{3}(f(Q,B)-Qf_{Q}-Bf_{B})-6a\dot{a}^{2}f_{Q}-6a^{2}\dot{a}(f_{BQ}\dot{Q}+f_{BB}\dot{B}) - \rho_{m0}.
\end{equation}
The equations of motion for a dynamical system can be obtained from the Euler-Lagrange equation,
\begin{equation}\label{Eq:Euler formula}
\frac{d}{dt}\left( \frac{\partial \mathcal{L}}{\partial \dot{q_{i}}} \right)- \frac{\partial \mathcal{L}}{\partial {q_{i}}}=0.
\end{equation}

In this scenario, we analyze situations where $q_{i}$ assumes the values $a$, $Q$, and $B$, respectively. In this context, $q_{i}$ denotes the generalized coordinates of the configuration space $\lbrace Q, B \rbrace$. We can define the equations of motion in the following way
\begin{eqnarray}
\frac{d}{dt}\left( \frac{\partial \mathcal{L}}{\partial \dot{a}} \right)- \frac{\partial \mathcal{L}}{\partial {a}}=0\label{Eq:Euler formula_a}\,,\\
\frac{d}{dt}\left( \frac{\partial \mathcal{L}}{\partial \dot{Q}} \right)- \frac{\partial \mathcal{L}}{\partial {Q}}=0 \label{Eq:Euler formula_Q}\,,\\
\frac{d}{dt}\left( \frac{\partial \mathcal{L}}{\partial \dot{B}} \right)- \frac{\partial \mathcal{L}}{\partial {B}}=0\label{Eq:Euler formula_B}\,.
\end{eqnarray}

From Eq.~ \eqref{Eq:main_Lagrangian} and \eqref{Eq:Euler formula_a}, we have derived an equation that corresponds to Eq. \eqref{Eq:Pde}, indicating that the Lagrangian $ \mathcal{L} $ aids in recovering the modified Friedmann equation.
The Euler Lagrange equations associated with the auxiliary variables $Q$ and $B$ provide a useful consistency check of the point-like Lagrangian. Starting from Eq.~\eqref{Eq:main_Lagrangian}, variation with respect to $Q$ yields
\begin{equation}
f_{QQ}\left(Q+6H^2\right) + f_{QB} \left[ B-6\left( \frac{\ddot a}{a}+2H^2 \right) \right] = 0,
\label{eq:EL_Q_constraint}
\end{equation}
Similarly, variation with respect to $B$ gives
\begin{equation}
f_{QB}\left(Q+6H^2\right) + f_{BB} \left[ B-6\left( \frac{\ddot a}{a}+2H^2 \right) \right] =0.
\label{eq:EL_B_constraint}
\end{equation}

Eqs.~\eqref{eq:EL_Q_constraint} and \eqref{eq:EL_B_constraint} can be written in matrix form as
\begin{equation}
\begin{pmatrix}
f_{QQ} & f_{QB} \\ f_{QB} & f_{BB}
\end{pmatrix}
\begin{pmatrix}
Q+6H^2 \\  B-6\left(\dfrac{\ddot a}{a}+2H^2\right)
\end{pmatrix} = \begin{pmatrix} 0\\ 0
\end{pmatrix}.
\label{eq:Hessian_constraint}
\end{equation}
The coefficient matrix in Eq.~\eqref{eq:Hessian_constraint} is the Hessian matrix of $f(Q,B)$ with respect to $Q$ and $B$. For a non-degenerate theory, its determinant satisfies
\begin{equation}
\Delta_f \equiv f_{QQ}f_{BB}-f_{QB}^{,2}
\neq 0.
\label{eq:Hessian_nonzero}
\end{equation}
Under this condition, the Hessian matrix is invertible, and the homogeneous system \eqref{eq:Hessian_constraint} admits only the trivial solution. Consequently,
\begin{equation}
Q+6H^2=0, \qquad B-6\left( \frac{\ddot a}{a}+2H^2 \right)=0.
\end{equation}

Therefore, provided that the Hessian of $f(Q,B)$ is non-singular, the Euler Lagrange equations for the auxiliary variables $Q$ and $B$ reproduce the defining FLRW relations for the non-metricity scalar and the boundary term. This establishes the dynamical consistency of the point-like Lagrangian with the geometrical constraints imposed in the original action. If $\Delta_f=0$, the Hessian is singular and the two Euler Lagrange equations are not independent; such degenerate cases must therefore be examined separately.

For Lagrangian $ \mathcal{L} $, the total energy (Hamiltonian) is given by
\begin{equation}\label{EQ:Hamilton_formula}
E_{\mathcal{H}}(a,\dot{a},Q,\dot{Q},B,\dot{B})=\frac{\partial \mathcal{L}}{\partial \dot{a}} \dot{a}+\frac{\partial \mathcal{L}}{\partial \dot{Q}} \dot{Q}+\frac{\partial \mathcal{L}}{\partial \dot{B}} \dot{B}-\mathcal{L}\,.
\end{equation}

Based on Eq.~\eqref{Eq:main_Lagrangian} and \eqref{EQ:Hamilton_formula}, and by setting $ E_{\mathcal{H}}=0 $, we derived an equation that is consistent with Eq. \eqref{Eq:rhode}. We have shown that the Lagrangian, defined as point-like and presented in Eq.~\eqref{Eq:main_Lagrangian}, can generate all the necessary equations of motion. This validates that it is the intended Lagrangian.

%%%%%%%%%%%%%%%%%%%%%%%%%%%%%%%%%%%%%%%%%%%%%%%%%%%%%%%%%%%%%%%%%%%%%%%%%%%%%%%%%%%%%%%%%%%%%%%%%%%%%%%%%%%%%%

%%%%%%%%%%%%%%%%%%%%%%%%%%%%%%%%%%%%%%%%%%%%%%%%%%%%%%%%%%%%%%%%%%%%%%%%%%%%%%%%%%%%%%%%%%%%%
\section{Noether Symmetries}\label{SEC:Noether Symmetry}

Noether symmetries are essential in physics, especially for examining the integrability of differential equations and streamlining intricate dynamical systems. Their presence is frequently associated with conserved quantities of physical significance. In cosmology, utilizing the Noether symmetry method is an effective means of obtaining exact solutions to the fundamental equations.

To demonstrate this, let us examine a system defined by $ n $ generalized coordinates $ x_{j} $ along with an independent variable $ t $, wherein the dynamics are described by a Lagrangian $ \mathcal{L} $. An infinitesimal one-parameter point transformation can be written as:
\begin{equation}
\bar{t} = \psi(t, x_{k}, \epsilon), \quad \bar{x}^{A} = \eta(t, x_{k}, \epsilon)\,,
\end{equation}
where $ \psi $ and $ \eta $ define the transformation. The generator for such a transformation can be written as
\begin{equation}
\mathcal{X} = \xi(t, x_{k}, \epsilon) \frac{\partial}{\partial t} + \alpha_{j}(t, x_{k}, \epsilon) \frac{\partial}{\partial x_{k}},
\end{equation}
with the coefficients $ \xi(t, x_{k}) $ and $ \alpha_{j}(t, x_{k}) $ determined by:
\begin{equation}
\xi(t, x_{k}) = \frac{\partial \psi(t, x_{k}, \epsilon)}{\partial \epsilon} \bigg|_{\epsilon \to 0}, \quad \quad \quad \quad \quad  \alpha_{j}(t, x_{k}) = \frac{\partial \eta(t, x_{k}, \epsilon)}{\partial \epsilon} \bigg|_{\epsilon \to 0}\,.
\end{equation}
The $ n^{\text{th}} $ prolongation of the generator vector is defined as \cite{Dialektopoulos_2022} 
\begin{equation}
\mathcal{X}^{[n]} = \mathcal{X} + \alpha_{j}^{[1]} \frac{\partial}{\partial \dot{x}_{j}} + \dots + \alpha_{j}^{[n]} \frac{\partial}{\partial x_{j}^{(n)}}\,,
\end{equation}
The coefficients $ \alpha_{j}^{[1]} $ and $ \alpha_{j}^{[n]} $ are derived through a recursive process.
\begin{equation}
\alpha_{j}^{[1]} = \frac{d\alpha_{j}}{dt} - \dot{x}_{j} \frac{d\xi}{dt}, \quad \quad \quad \quad \quad \quad\alpha_{j}^{[n]} = \frac{d\alpha_{j}^{[n-1]}}{dt} - x_{j}^{(n)} \frac{d\xi}{dt}\,.
\end{equation}

For the Euler-Lagrange equations of a dynamical system with Lagrangian $ \mathcal{L} = \mathcal{L}(t, x_{j}, \dot{x}_{j}) $ to exhibit invariance under the transformation, a function $ g = g(t, x_{j}) $ must exist so that the Rund-Trautman identity is fulfilled
\begin{equation}\label{eq:noether_condition}
\mathcal{X}^{[1]}\mathcal{L} + \mathcal{L} \frac{d\xi(t, x_{j})}{dt} = \frac{dg(t, x_{j})}{dt}\,,
\end{equation}

where \( \mathcal{X}^{[1]} \) represents the first prolongation of \( \mathcal{X} \). If the generator \( \mathcal{X} \) meets this requirement, it is recognized as a Noether symmetry for the system defined by \( \mathcal{L} \). Based on Noether’s theorem, the existence of a symmetry guarantees that there is a conserved quantity, known as the Noether charge, which is defined by
\begin{equation}\label{noether_charge}
\mathcal{Q}_{0} = \sum_{j} \alpha_{j} \frac{\partial \mathcal{L}}{\partial \dot{q}_{j}} = \text{constant}\,.
\end{equation}

In this context, \( q_{j} \) denotes the coordinates within the configuration space, while \( \alpha_{j} \) signifies the components of the Noether factors. This conserved quantity offers essential understanding of the system's dynamics.

%%%%%%%%%%%%%%%%%%%%%%%%%%
%%%%%%%%%%%%%%%%%%%%%%%%%%%%%%%%%%%%%%%%%%%%%
\subsection{Noether symmetries in $f(Q,B)$ gravity}  \label{Sec:NoetherfQB} 

This section explores Noether symmetry in the context of $f(Q, B)$ theory. The Noether symmetry approach \cite{Capozziello_1993} is employed to identify potential symmetries for the Lagrangian dynamical system ~\eqref{Eq:main_Lagrangian}. Noether symmetry is a valuable tool for discovering exact solutions associated with a given Lagrangian, with the resulting models being fundamentally justified. The Noether symmetry generator is represented by a vector $\mathcal{X}$. The existence of symmetry relies on a vector defined within the tangent space of the Lagrangian $\mathcal{L}$. This section focuses on a one-parameter point transformation in the configuration space $(t, a, Q, B)$. The generator is expressed as
\begin{equation}\label{Eq:noethervectorheberal}
\chi = \xi(t,a,Q,B)\frac{\partial}{\partial t} + \theta_1(t,a,Q,B)\frac{\partial}{\partial a}
+ \theta_2(t,a,Q,B)\frac{\partial}{\partial Q} + \theta_3(t,a,Q,B)\frac{\partial}{\partial B},
\end{equation}
Here, $\xi$ $ \theta_1$, $ \theta_2$ and $ \theta_3$ are functions of the generalised coordinates $ t $, $ a $, $ Q $ and $ B $. The first prolongation of the generator vector is
\begin{equation}
\chi^{[1]} = \chi + \theta_1^{[1]}\frac{\partial}{\partial \dot{a}} + \theta_2^{[1]}\frac{\partial}{\partial \dot{Q}} + \theta_3^{[1]}\frac{\partial}{\partial \dot{B}},
\end{equation}
with
\begin{equation}
\theta_1^{[1]} = \frac{\partial \theta_1}{\partial t} + \dot{a}\frac{\partial \theta_1}{\partial a}
+ \dot{Q}\frac{\partial \theta_1}{\partial Q} + \dot{B}\frac{\partial \theta_1}{\partial B} - \dot{a}\frac{\partial \xi}{\partial t} - \dot{a}^{\,2}\frac{\partial \xi}{\partial a}
- \dot{a}\dot{Q}\frac{\partial \xi}{\partial Q}\ - \dot{a}\dot{B}\frac{\partial \xi}{\partial B},
\end{equation}
\begin{equation}
\theta_2^{[1]} = \frac{\partial \theta_2}{\partial t} + \dot{a}\frac{\partial \theta_2}{\partial a}
+ \dot{Q}\frac{\partial \theta_2}{\partial Q} + \dot{B}\frac{\partial \theta_2}{\partial B} - \dot{Q}\frac{\partial \xi}{\partial t} - \dot{a}\dot{Q}\frac{\partial \xi}{\partial a}
- \dot{Q}^{\,2}\frac{\partial \xi}{\partial Q} - \dot{B}\dot{Q}\frac{\partial \xi}{\partial B}.
\end{equation}
\begin{equation}
\theta_3^{[1]} = \frac{\partial \theta_3}{\partial t} + \dot{a}\frac{\partial \theta_3}{\partial a}
+ \dot{Q}\frac{\partial \theta_3}{\partial Q} + \dot{B}\frac{\partial \theta_3}{\partial B} - \dot{B}\frac{\partial \xi}{\partial t} - \dot{a}\dot{B}\frac{\partial \xi}{\partial a}
- \dot{Q}\dot{B}\frac{\partial \xi}{\partial Q} - \dot{B}^{\,2}\frac{\partial \xi}{\partial B}.
\end{equation}

The first term of Eq.~\eqref{eq:noether_condition}, $\chi^{[1]} \mathcal{L} $, is
\begin{equation}
 \begin{aligned}
\chi^{[1]} \mathcal{L}= & 3 \theta_{1} a^{2}\left(f-Q f_{Q}-B f_{B}\right)-\theta_{2}\left(Q f_{Q Q}+B f_{B Q}\right)-\theta_{3}\left(Q f_{Q B}+B f_{B B}\right) + 12a\dot{a}^{3} f_{Q} \frac{\partial \xi }{\partial a} \\
 & + \dot{a}^{2}\left[-6 \theta_{1} f_{Q} -12 a f_{Q} \frac{\partial \theta_{1}}{\partial a} \right.\left. -6 a^{2} f_{B Q} \frac{\partial \theta_{2}}{\partial a} -6 a^{2} f_{B B} \frac{\partial \theta_{3}}{\partial a} + 12 a f_{Q} \frac{\partial \xi}{\partial t} + 12 a f_{Q} \dot{Q} \frac{\partial \xi}{\partial Q} \right.\\
 &\left.+ 12 a f_{Q} \dot{B} \frac{\partial \xi}{\partial B} + 12 a^{2} f_{B Q} \dot{Q} \frac{\partial \xi}{\partial a} + 12 a^{2} f_{B B} \dot{B} \frac{\partial \xi}{\partial a} \right] \\
& +\dot{Q}^{2}\left[-6 a^{2} f_{B Q} \frac{\partial \theta_{1}}{\partial Q}+12 \dot{a} a^{2} f_{B Q} \frac{\partial \xi}{\partial Q}\right]+\dot{B}^{2}\left[-6 a^{2} f_{B B} \frac{\partial \theta_{1}}{\partial B} +12 a a^{2} f_{B B} \frac{\partial \xi}{\partial B}\right] \\
& +\dot{Q} \dot{B}\left[-6 a^{2} f_{B B} \frac{\partial \theta_{1}}{\partial Q}-6 a^{2} f_{B Q} \frac{\partial \theta_{1}}{\partial B}+12 \dot{a} a^{2} f_{B B} \frac{\partial \xi}{\partial Q}+12 \dot{a} a^{2} f_{B Q} \frac{\partial \xi}{\partial B}\right] \\
& +\dot{a} \dot{Q}\left[-12 a f_{B Q} -12 a f_{Q} \frac{\partial \theta_{1}}{\partial Q} -6 a^{2} f_{B Q} \frac{\partial \theta_{1}}{\partial a}-6 a^{2} f_{B Q} \frac{\partial \theta_{2}}{\partial Q}-6 a^{2} f_{B B} \frac{\partial \theta_{2}}{\partial Q}+12 a^{2} f_{B Q} \frac{\partial \xi}{\partial t}\right] \\
& +\dot{a} \dot{B}\left[-12 a f_{B B} -12 a f_{Q} \frac{\partial \theta_{1}}{\partial B} -6 a^{2} f_{B B} \frac{\partial \theta_{1}}{\partial a}-6 a^{2} f_{B Q} \frac{\partial \theta_{2}}{\partial B}-6 a^{2} f_{B B} \frac{\partial \theta_{3}}{\partial B}+12 a^{2} f_{B B} \frac{\partial \xi}{\partial t}\right] \\
& + \dot{a} \left[ -12a f_{Q} \frac{\partial \theta_{1}}{\partial t} - 6 a^{2} f_{B Q} \frac{\partial \theta_{2}}{\partial t} - 6 a^{2} f_{B B} \frac{\partial \theta_{3}}{\partial t} \right] + \dot{Q} \left[ - 6 a^{2} f_{B Q} \frac{\partial \theta_{1}}{\partial t} \right] + \dot{B} \left[ - 6 a^{2} f_{B B} \frac{\partial \theta_{1}}{\partial t} \right]
\end{aligned}
\end{equation}
The second term of Eq.\eqref{eq:noether_condition}, $\mathcal{L}\dot{\xi}$, is
\begin{equation}
\begin{aligned}
\mathcal{L}\dot{\xi} = & \left( a^{3}(f-Qf_{Q}-Bf_{B})-6a\dot{a}^{2}f_{Q}-6a^{2}\dot{a}(f_{BQ}\dot{Q}+f_{BB}\dot{B}) \right)\frac{\partial \xi}{\partial t}
\\
& + \left( \dot{a}a^{3}(f-Qf_{Q}-Bf_{B})-6a\dot{a}^{3}f_{Q}-6a^{2}\dot{a}^{2}(f_{BQ}\dot{Q}+f_{BB}\dot{B}) \right)\frac{\partial \xi}{\partial a}
\\
& + \left( \dot{Q} a^{3}(f-Qf_{Q}-Bf_{B})-6a\dot{a}^{2} \dot{Q} f_{Q}-6a^{2}\dot{a} \dot{Q}(f_{BQ}\dot{Q}+f_{BB}\dot{B}) \right)\frac{\partial \xi}{\partial Q}
\\
& + \left( \dot{B} a^{3}(f-Qf_{Q}-Bf_{B})-6a\dot{a}^{2} \dot{B} f_{Q}-6a^{2}\dot{a} \dot{B}(f_{BQ}\dot{Q}+f_{BB}\dot{B}) \right)\frac{\partial \xi}{\partial B}.
\end{aligned}
\end{equation}
Furthermore, the right-hand side of Eq.~\eqref{eq:noether_condition} is
\begin{equation}
\dot{g} = \frac{\partial g}{\partial t} +\dot{a}\frac{\partial g}{\partial a}
+\dot{Q}\frac{\partial g}{\partial Q} +\dot{B}\frac{\partial g}{\partial B}
\end{equation}

According to the well-established Noether theorem, a constant of motion (Noether charge) will exist \cite{Capozziello_1993}. This constant can be identified as:
\begin{equation}\label{Eq:noethercharge}
\mathcal{Q}_{0}=\sum_{i}\theta_{i}\frac{\partial \mathcal{L}}{\partial\dot{q_{i}}}=\theta_1 \frac{\partial \mathcal{L}}{\partial \dot{a}}+\theta_2 \frac{\partial \mathcal{L}}{\partial \dot{Q}}+\theta_3 \frac{\partial \mathcal{L}}{\partial \dot{B}}=constant
\end{equation}
The explicit evaluation yields a second-degree expression in $ \dot{a} $, $ \dot{Q} $ and $ \dot{B} $, with coefficients that depend solely on $ a $, $ Q $ and $ B $. Consequently, each of them should be set to zero individually \cite{Capozziello_1993}. Therefore, by equating the coefficients of $\dot{a}^{2}$, $\dot{Q}^{2}$, $\dot{B}^{2}$, $\dot{a}\dot{Q}$, $\dot{a}\dot{B}$ and $\dot{B}\dot{Q}$ to zero, we get 

\begin{align}
3\theta_{1}a^{2}
\left(f-Qf_{Q}-Bf_{B}\right)
&-\theta_{2}\left(Qf_{QQ}+Bf_{BQ}\right)
-\theta_{3}\left(Qf_{QB}+Bf_{BB}\right)
\nonumber\\
&+a^{3}\left(f-Qf_{Q}-Bf_{B}\right)
\frac{\partial\xi}{\partial t}
=\frac{\partial g}{\partial t},
\label{Eq:system1}
\\[1ex]
-12af_{Q}\frac{\partial\theta_{1}}{\partial t}
&-6a^{2}f_{BQ}\frac{\partial\theta_{2}}{\partial t}
-6a^{2}f_{BB}\frac{\partial\theta_{3}}{\partial t}
\nonumber\\
&+a^{3}\left(f-Qf_{Q}-Bf_{B}\right)
\frac{\partial\xi}{\partial a}
=\frac{\partial g}{\partial a},
\label{Eq:system2}
\\[1ex]
-6a^{2}f_{BQ}\frac{\partial\theta_{1}}{\partial t}
&+a^{3}\left(f-Qf_{Q}-Bf_{B}\right)
\frac{\partial\xi}{\partial Q}
=\frac{\partial g}{\partial Q},
\label{Eq:system3}
\\[1ex]
-6a^{2}f_{BB}\frac{\partial\theta_{1}}{\partial t}
&+a^{3}\left(f-Qf_{Q}-Bf_{B}\right)
\frac{\partial\xi}{\partial B}
=\frac{\partial g}{\partial B},
\label{Eq:system4}
\\[1ex]
-6\theta_{1}f_{Q}
&-12af_{Q}\frac{\partial\theta_{1}}{\partial a}
-6a^{2}f_{BQ}\frac{\partial\theta_{2}}{\partial a}
-6a^{2}f_{BB}\frac{\partial\theta_{3}}{\partial a}
\nonumber\\
&+6af_{Q}\frac{\partial\xi}{\partial t}
+6af_{Q}\dot{Q}\frac{\partial\xi}{\partial Q}
+6af_{Q}\dot{B}\frac{\partial\xi}{\partial B}
\nonumber\\
&+6a^{2}f_{BQ}\dot{Q}\frac{\partial\xi}{\partial a}
+6a^{2}f_{BB}\dot{B}\frac{\partial\xi}{\partial a}
=0,
\label{Eq:system5}
\\[1ex]
-6a^{2}f_{BQ}\frac{\partial\theta_{1}}{\partial Q}
&=0,
\qquad
6\dot{a}a^{2}f_{BQ}\frac{\partial\xi}{\partial Q}=0,
\nonumber\\
-6a^{2}f_{BB}\frac{\partial\theta_{1}}{\partial B}
&=0,
\qquad
6a^{3}f_{BB}\frac{\partial\xi}{\partial B}=0,
\nonumber\\
12af_{Q}\frac{\partial\xi}{\partial a}
&=0,
\label{Eq:system6}
\\[1ex]
-12af_{BQ}
&-12af_{Q}\frac{\partial\theta_{1}}{\partial Q}
-6a^{2}f_{BQ}\frac{\partial\theta_{1}}{\partial a}
\nonumber\\
&-6a^{2}f_{BQ}\frac{\partial\theta_{2}}{\partial Q}
-6a^{2}f_{BB}\frac{\partial\theta_{2}}{\partial Q}
+6a^{2}f_{BQ}\frac{\partial\xi}{\partial t}
=0,
\label{Eq:system7}
\\[1ex]
-12af_{BB}
&-12af_{Q}\frac{\partial\theta_{1}}{\partial B}
-6a^{2}f_{BB}\frac{\partial\theta_{1}}{\partial a}
\nonumber\\
&-6a^{2}f_{BQ}\frac{\partial\theta_{2}}{\partial B}
-6a^{2}f_{BB}\frac{\partial\theta_{3}}{\partial B}
+6a^{2}f_{BB}\frac{\partial\xi}{\partial t}
=0.
\label{Eq:system8}
\end{align}

In this case, the function $f(Q,B)$, $\xi(t,a,Q,B)$, $\theta_1(t,a,Q,B)$, $\theta_2(t,a,Q,B)$, and $\theta_3(t,a,\\Q,B)$ are the unknown variables. It is possible to declare the existence of Noether symmetry if at least one of the variables is not zero. There are two ways to work it out and identify symmetry. The unknown variables and functions can be found by directly solving the system of partial differential Eq.~(\ref{Eq:system1}--\ref{Eq:system8}). A second approach allows one to impose specific forms of $f(Q,B)$ and find corresponding symmetries. From a physical viewpoint, the second approach is better because it allows to study reliable models. By the first strategy, solutions can be achieved but, in most cases, they are implicit functions that do not allow a physical analysis.

From Eq.~\eqref{Eq:system6}, we can see that the Noether Coefficient $\theta_1 $ is independent of $Q$ and $B$, i.e $\theta_1=\theta_1(t,a)$, while the temporal component $\xi$ is independent of the variables $a$, $Q$, and $B$, i.e $\xi=\xi(t)$. To address the symmetries in $ f(Q, B) $ cosmology, we need to consider some functional form of $f(Q, B)$. In the next section, we consider the power-law form of $f(Q, B)$ and study its cosmological aspects.

\subsection{Power-Law $f(Q,B)$ Model} \label{Sec:modelnoether}

We now specialize the Noether symmetry analysis to a multiplicative power-law form of the gravitational function. This class of models is particularly convenient for investigating scaling symmetries because the dependence on the non-metricity scalar $Q$ and the boundary term $B$ is homogeneous. We consider
\begin{equation}
f(Q,B)=f_0(-Q)^{\alpha}B^{\beta},
\label{eq:powerlaw_model}
\end{equation}
where $f_0$, $\alpha$, and $\beta$ are constants. For $Q\neq0$ and $B\neq0$, the first derivatives of Eq.~\eqref{eq:powerlaw_model} are
\begin{equation}
f_Q=\frac{\alpha f}{Q}, \qquad f_B=\frac{\beta f}{B},
\label{eq:powerlaw_first_derivatives}
\end{equation}
while the relevant second derivatives are
\begin{equation}
f_{QQ}=\frac{\alpha(\alpha-1)f}{Q^2}, \qquad f_{QB}=\frac{\alpha\beta f}{QB}, \qquad
f_{BB}=\frac{\beta(\beta-1)f}{B^2}.
\label{eq:powerlaw_second_derivatives}
\end{equation}
It follows immediately that
\begin{equation}
f-Qf_Q-Bf_B =  (1-\alpha-\beta)f.
\label{eq:powerlaw_homogeneity}
\end{equation}

Before applying the Noether symmetry condition, it is useful to examine the non-degeneracy of the minisuperspace formulation. As shown by the Euler Lagrange equations associated with the auxiliary variables $Q$ and $B$, the geometrical constraints can be expressed as Eq.~\eqref{eq:Hessian_constraint}. Therefore, the standard geometrical definitions are recovered independently whenever the Hessian of $f(Q,B)$ is non-singular. For the model \eqref{eq:powerlaw_model}, its determinant is
\begin{align}
\Delta_f \equiv f_{QQ}f_{BB}-f_{QB}^{,2} = \frac{\alpha\beta f^2}{Q^2B^2} \left[ (\alpha-1)(\beta-1)-\alpha\beta \right] = \frac{\alpha\beta(1-\alpha-\beta)f^2} {Q^2B^2}. 
\label{eq:powerlaw_hessian}
\end{align}
Thus, for $Q\neq0$, $B\neq0$, and $f\neq0$, the generic non-degenerate branch is characterized by
\begin{equation}
\alpha\neq0,\qquad \beta\neq0,\qquad \alpha+\beta\neq1.
\label{eq:powerlaw_nondegenerate}
\end{equation}

The cases $\alpha=0$, $\beta=0$, or $\alpha+\beta=1$ correspond to degenerate branches for which the two auxiliary-field equations are not independent. They should therefore be treated separately and should not be obtained merely by taking limits of results derived under the non-degeneracy condition \eqref{eq:powerlaw_nondegenerate}.
Using Eqs.~(\ref{eq:powerlaw_first_derivatives}--\ref{eq:powerlaw_homogeneity}), the gravitational part of the point-like Lagrangian \eqref{Eq:main_Lagrangian} becomes
\begin{align}
\mathcal{L} = a^3(1-\alpha-\beta)f -6a\frac{\alpha f}{Q}\dot a^2 -6a^2\dot a \left[ \frac{\alpha\beta f}{QB}\dot Q + \frac{\beta(\beta-1)f}{B^2}\dot B \right] -\rho_{m0},
\label{eq:powerlaw_lagrangian}
\end{align}

We substitute the power-law expression of \( f(Q, B) \) into the system of partial differential Eqs.~ (\ref{Eq:system1}-\ref{Eq:system8}), and by applying the method of separation of variables, we can derive the Noether coefficients for the Noether vector (\ref{Eq:noethervectorheberal}). For the general power-law model, we have obtained the following Noether coefficients
\begin{align}
\xi & = \xi_0t+\xi_1, \label{eq:powerlaw_xi}\ \\
\theta_1 & = \frac{\xi_0}{3} \left(2\alpha+2\beta-1\right)a, \label{eq:powerlaw_theta1} \\
\theta_2 & = -2\xi_0Q, \label{eq:powerlaw_theta2}\ \\
\theta_3 & = -2\xi_0B, \label{eq:powerlaw_theta3}\ \\
g & = g_0, \label{eq:powerlaw_gauge}\
\end{align}
where $\xi_0$, $\xi_1$, and $g_0$ are constants. Consequently, the Noether vector can be written as
\begin{equation}
\chi = (\xi_0t+\xi_1)\frac{\partial}{\partial t} + \frac{\xi_0}{3} (2\alpha+2\beta-1)a \frac{\partial}{\partial a} -2\xi_0Q\frac{\partial}{\partial Q} -2\xi_0B\frac{\partial}{\partial B}.
\label{eq:powerlaw_scaling_generator}
\end{equation}
The constant $\xi_1$ generates time translations, whereas the part proportional to $\xi_0$ corresponds to a simultaneous scaling transformation of $t$, $a$, $Q$, and $B$. The total energy ($E_{\mathcal{H}}$) \eqref{EQ:Hamilton_formula} associated with $\mathcal{L}$ is therefore
\begin{align}
E_{\mathcal{H}}  = -a^3(f-Qf_Q-Bf_B) -6af_Q\dot a^2 -6a^2\dot a \left( f_{QB}\dot Q+f_{BB}\dot B \right) +\rho_{m0}.
\label{eq:correct_energy_function}
\end{align}
Using the Noether charge \eqref{noether_charge}, this reduces to
\begin{eqnarray}
&&\frac{a\xi_0}{3} (2\alpha+2\beta-1) (-12a\dot a f_Q -6a^2 \left( f_{QB}\dot Q+f_{BB}\dot B \right)) \nonumber \\&&-2\xi_0Q (-6a^2\dot a f_{QB}) -2\xi_0B (-6a^2\dot a f_{BB}) = \mathrm{constant}.
\label{eq:scaling_first_integral}
\end{eqnarray}
Motivated by the scaling character of the Noether symmetry, we consider the following form of scale factor
\begin{equation}
a(t)=a_0(t-t_0)^s,
\label{eq:powerlaw_scale_factor_ansatz}
\end{equation}

where $a_0>0$, $t_0$, and $s$ are constants. we have consider $\tau=t-t_0$. For the power-law scale-factor solution, we obtain the Hubble parameter $H=\frac{s}{\tau}$. The non-metricity $Q$ and boundary term $B$ can be written as 
\begin{equation}
Q=-\frac{6s^2}{\tau^2}, \qquad B=\frac{6s(3s-1)}{\tau^2}
\label{eq:powerlaw_QB}
\end{equation}

Since both \( Q \) and \( B \) scale as \( \tau^{-2} \), the gravitational function behaves as \( f(Q, B) \propto \tau^{-2(\alpha + \beta)} \). For the scale-factor solution and considered power-law \( f(Q, B) \) model, the scale-factor Euler-Lagrange Eq.~\eqref{Eq:Euler formula_a} gives, 
\begin{equation}
\left(3s-2\alpha-2\beta\right) \left[ 2(\alpha+\beta-1)(3s-1) + 2\beta(\alpha+\beta-1) \right] =0.
\label{eq:factorized_scale_equation}
\end{equation}
For the generic non-degenerate conditions defined in Eq.~\eqref{eq:powerlaw_nondegenerate} for the power-law $f(Q, B)$ model, we have $\alpha \neq 0$, $\beta \neq 0$, and $\alpha +\beta \neq 1$. Under these non-degeneracy conditions, the second factor of Eq.~\eqref{eq:factorized_scale_equation} is non-zero, i.e., $(2(\alpha+\beta-1)(3s-1) + 2\beta(\alpha+\beta-1)) \neq 0$. Therefore, the first factor must vanish $(3s-2\alpha-2\beta=0)$. Thus, we obtain 
\begin{equation}
s = \frac{2(\alpha+\beta)}{3} .
\label{eq:powerlaw_exponent}
\end{equation}
The scale factor solution becomes
\begin{equation}
a(t) = a_0(t-t_0)^{\frac{2(\alpha+\beta)}{3}}.
\label{eq:exact_powerlaw_scale_factor}
\end{equation}
The Hubble parameter, non-metricity scalar $Q$ and boundary term $B$ can be defined as 
\begin{eqnarray}
H(t) = \frac{2(\alpha+\beta)} {3(t-t_0)}, \quad \quad  Q(t) = -\frac{8(\alpha+\beta)^2} {3(t-t_0)^2}, \qquad B(t) = \frac{ 4(\alpha+\beta) \left[2(\alpha+\beta)-1\right] } {(t-t_0)^2}.   
\end{eqnarray}

We observe that various cosmologically significant scenarios can be derived from the precise power-law solution. A radiation-dominated solution  is for
\begin{equation}
 a(t)= a_0 (t-t_0)^{\frac{1}{2}},\quad \quad \text{for model parameter condition}, \quad \quad \alpha+\beta=\frac{3}{4},    
\end{equation}
A matter-dominated solution is for 
\begin{equation}
 a(t)= a_0 (t-t_0)^{\frac{2}{3}},\quad \quad \text{for model parameter condition}, \quad \quad \alpha+\beta=1,    
\end{equation}
A stiff-matter solution is for 
\begin{equation}
 a(t)= a_0 (t-t_0)^{\frac{1}{3}},\quad \quad \text{for model parameter condition}, \quad \quad \alpha+\beta=\frac{1}{2},    
\end{equation}
The deceleration parameter is given by
\begin{equation}
q = -1-\frac{\dot H}{H^2} = \frac{3}{2(\alpha+\beta)}-1.
\label{eq:deceleration_parameter}
\end{equation}
Thus, the model describes accelerated expansion phase ($q<0$) of the Universe for the condition $\alpha+\beta>\frac{3}{2}$. We stress that the exact solution \eqref{eq:exact_powerlaw_scale_factor} is obtained for the vacuum (or geometry-dominated) dynamics, i.e., in the absence of the dust contribution $\rho_{m0}$ in the Lagrangian \eqref{Eq:main_Lagrangian}; a pure power-law scale factor cannot solve the full system once pressureless matter is included. The exact solution should therefore be understood as describing the asymptotic regimes in which the geometric sector dominates the expansion — in particular, the late-time accelerated epoch for $\alpha+\beta>\frac{3}{2}$.

%%%%%%%%%%%%%%%%%%%%%%%%%%%%%%%%%%%%%%%%%%%%%%%%%
\section{Cosmological observations $f(Q, B)$ model}\label{SEC:cosmologicalobsevation}

In the previous section, we studied the Noether symmetry approach in the power-law $f(Q, B)$ gravity model. We have also obtained the exact solution of the field equation. We found that the model shows the accelerating phase of the Universe for the conditions $\alpha + \beta >\frac{3}{2}$. To motivate their solution, we will study the assumed model using the cosmological observational data set in this section to assess its viability. For this, we will consider the Cosmic Chronometer (CC), Type Ia Supernovae (SNe Ia) data, and the recently introduced BAO DESI DR2 data set. For this $f(Q, B)$ model, we have three free parameters $f_0$, $\alpha$, and $\beta$. So, we have calculated the value of model parameter $f_0$ from Eq. \eqref{first_field_equation}  at present time as  
\begin{equation}\label{f0_value}
 f_0 = \frac{6 H_0^{2}(1-\Omega_{m0}- \Omega_{r0})}{\mathcal{D}} \,.  
\end{equation}

\begin{eqnarray}
 &&\mathcal{D} = -(-Q_0)^{\alpha} B_0^{\beta}-12 H_0^2 \alpha (-Q_0)^{\alpha-1} B_0^{\beta}- 3 \beta (-Q_0)^{\alpha+1} B_0^{\beta-1}+ 6 \dot{H_0} \beta (-Q_0)^{\alpha} B_0^{\beta-1}\nonumber \\ &&+6 H_0 \beta \alpha (-Q_0)^{\alpha-1} \dot{Q_0} B_0^{\beta-1}-6 H_0 (-Q_0)^{\alpha} B_0^{\beta-2} \beta (\beta-1) \dot{B_0}  
\end{eqnarray}

The terms  $\Omega_{m0}$ and $ \Omega_{r0}$ are the density parameters for the matter and radiation sectors at present respectively. The representation of $f_0$ using $\alpha$, $\beta$, and other cosmological parameters is beneficial as it decreases the number of free parameters in the system, thereby simplifying the analysis. As a result, rather than incorporating three new parameters like in the original equation, only two model parameters are added, resulting in a more straightforward and refined model. The parameters $\alpha$ and $\beta$ can be determined by using MCMC analyses on observational data. Additionally, \( B_0 \) denotes the value of the boundary term at present, while \( Q_0 \) signifies the non-metricity value at that same moment. In particular, \( B_0 = B_{|t=t_{0}} \) and \( \dot{B}_{|t=t_{0}} \) represents its derivative assessed at the present time. Likewise, \( Q_0 = Q_{|t=t_{0}} \) and \( \dot{Q}_{|t=t_{0}} \) is its derivative, which is also considered at the current time.
To study the cosmological observations, we have defined the Hubble parameter to obtain the theoretical values. For this, we have obtained the following form of the Hubble parameter from  Eq.\eqref{first_field_equation} as  
\begin{eqnarray}\label{hubbleparameter_Equation}
 &&3 H^2(z)+18 f_0 \beta (\beta-1) (-Q)^{\alpha} B^{\beta-2} (1+z)^2 H^3(z) H''(z)= 3H_0^2 (1+z)^3 \Omega_{m0} + 3H_0^2 (1+z)^4 \Omega_{r0} \nonumber \\&&-\frac{1}{2} f_0 (-Q)^{\alpha} B^{\beta}- 6 H^2(z) \alpha  (-Q)^{\alpha-1} B^{\beta}+ 9 f_0 \beta (-Q)^{\alpha} B^{\beta-1} H^2(z) - 3 f_0 \beta (-Q)^{\alpha} B^{\beta-1} \nonumber \\&&(1+z) H(z) H'(z)+ 36 f_0 \alpha \beta (-Q)^{\alpha-1} B^{\beta-1} (1+z) H^3(z) H'(z) + 90 f_0 \beta (-Q)^{\alpha} (\beta-1) B^{\beta-2}\nonumber \\&& (1+z) H^3(z) H'(z)- 18 f_0 \beta (-Q)^{\alpha} (\beta-1) B^{\beta-2} (1+z)^2 H^2(z) H'^{2}(z)\,.
\end{eqnarray}

In the above equation, prime ($'$) indicates the derivative with respect to redshift ($z$). As Eq. \eqref{hubbleparameter_Equation} represents a second-order nonlinear differential equation, it is typically impossible to find an exact analytical solution. Therefore, we use the numerical method to solve the equation numerically and examine the model's cosmological evolution. In order to solve the second-order differential equation, we must utilize appropriate initial conditions. To ensure a physically viable cosmological evolution, we impose initial conditions that are consistent with the standard $\Lambda$CDM model at the present epoch $H_0$. Since Eq.~\eqref{hubbleparameter_Equation} is a second-order differential equation, two independent initial conditions are required for its numerical integration. We choose the first initial condition as $H(0)=H_0$, where $H_0$ is the present-day Hubble parameter. The second initial condition is taken as $H'(0)=\frac{H_0} {2}\left(3\Omega_{m0}+4\Omega_{r0}\right)$, which is obtained from the standard $\Lambda$CDM cosmology and provides the initial slope of the Hubble parameter at the present epoch. These initial conditions are then used to solve the Eq. \eqref{hubbleparameter_Equation} numerically. We have taken $\Omega_{r0}=4.183699 \times 10^{-5}$.

To analyze cosmological data, we used the publicly available \texttt{emcee} package (Ref. \cite{Foreman_Mackey_2013}) to perform an MCMC analysis that integrates the data sets with the $f(Q, B)$ power-law model. MCMC is a sampling method that produces samples from the model's posterior distribution. This technique is frequently employed in Bayesian statistics, where the posterior distribution is used to estimate the model's parameters. Additionally, MCMC can be used to assess the uncertainty associated with a model. This procedure necessitates a substantial sample size to ensure that the models generated accurately reflect the data. Consequently, for each parameter, we obtain one and two dimensional distributions: the one-dimensional distribution shows the posterior of the parameter, while the two-dimensional distribution shows the covariance between two parameters. To achieve this, we generated the 1$\sigma$ (68\%) and 2$\sigma$ (95\%) confidence regions using the \texttt{ChainConsumer package} \cite{Hinton2016}. A prominent unresolved issue in contemporary cosmology is the inconsistency between the Hubble constant \( H_0 \) derived from early-Universe observations and that obtained from late-time measurements within the $\Lambda$CDM model. To address the $H_0$ tension and investigate the late-time cosmic acceleration, we employed various data sets and combined them for the model. 

\textbf{Cosmic Chronometers (CC):} We employed 31 data points estimated using the CC method, as referenced in \cite{Zhang_2014hz, Jimenez_2003cmb, Moresco_2016hubb, Simon_2005prd, M_Moresco_2012JCAP, Daniel_Stern_2010jcap, Moresco_2015mnras}. Through this technique, we can directly interpret the Hubble function across different redshifts, reaching up to $ z \lesssim 2 $. The CC data is advantageous as it assesses the age difference between two passively evolving galaxies that formed at the same time but are separated by a slight redshift interval ($\frac{\Delta z}{\Delta t}$), making it more efficient than other approaches that aim to ascertain the absolute age of galaxies \cite{Jimenez_2002}. The Hubble dataset is associated with star ages, which are derived from reliable stellar population synthesis models \cite{G_mez_Valent_2018jcap, L_pez_Corredoira_2017aa}, without relying on a cosmological model or the Cepheid distance scale. The corresponding estimate of $\chi^{2}_{CC}$ is provided by
\begin{equation}\label{chisqure_hz}
\chi^2_{CC}= \Delta H(z_i, \Theta)^{T} \,C_{CC}^{-1} \,\Delta H(z_i, \Theta) \,,   
\end{equation}
where $\Delta H(z_i, \Theta)= H(z_i, \Theta)-H_{\text{obs}}(z_i)$ and $C_{CC}^{-1}$ denotes the covariance matrix referenced in \cite{Moresco_2020covariance}. The expression $H(z_i, \Theta)$ signifies the theoretical Hubble parameter values corresponding to a given redshift $z_i$, whereas $H_{\text{obs}}(z_i)$ represents the observed Hubble parameter values at that same redshift $z_i$.

\textbf{Type Ia Supernovae data set :} The PN$^{+}\&$ SH0ES collection comprises 1,701 light curves \cite{Brout_2022panplus, Riess_2022panplus, Scolnic_2022panplus} sourced from 1,550 spectroscopically confirmed Type Ia supernovae (SNe Ia). These datasets will contribute to determining cosmological parameters in the Pantheon+ supernova project and the SH0ES distance-ladder investigation. The relative luminosity distance observations cover a redshift range of \(0.01 < z < 2.3\). The uniform, intrinsic brightness of these supernovae makes them significant for cosmological research, as they allow us to measure distances to distant galaxies by serving as standard candles. Specifically, the distance modulus function is defined as the difference between the observed apparent magnitude \( m \) and its absolute magnitude \( M \). At a redshift \( z_i \), the distance modulus function \( \mu(z_i, \Theta) \) can be formulated as,
\begin{equation}\label{modulus_function}
\mu(z_i, \Theta) = m - M = 5 \log_{10} \left[ D_L(z_i, \Theta) \right] + 25\,,    
\end{equation}
the luminosity distance $D_L(z_i, \Theta)$ defined as 
\begin{equation}\label{luminosity_distance}
D_L(z_i, \Theta) = c(1 + z_i) \int_0^{z_i} \frac{dz'}{H(z', \Theta)}\,.  
\end{equation}

Here, $c$ denotes the speed of light, $H(z,\Theta)$ represents the Hubble expansion rate, and $\Theta$ denotes the set of cosmological model parameters. Since the SNe Ia observations primarily constrain relative luminosity distances, the absolute magnitude $M$ is degenerate with the Hubble constant $H_0$ and can therefore be treated as a nuisance parameter in the absence of an external calibration. To calibrate the absolute distance scale, we incorporate the SH0ES distance-ladder information, which uses Cepheid variables in SNe Ia host galaxies to provide an independent calibration of the absolute magnitude of SNe Ia \cite{Scolnic_2022panplus, Riess_2022panplus}. The inclusion of the SH0ES calibration is therefore motivated by the need to break the $M- H_0$ degeneracy and to anchor the PN$^{+}$ SNe Ia Hubble diagram to an absolute distance scale. This allows the combined PN$^{+}$ \&SH0ES data to provide direct sensitivity to the present-day expansion rate $H_0$, while PN$^{+}$ supplies the relative distance-redshift relation over a broad redshift range. The combination is particularly relevant for investigating the late-time expansion of the Universe and the existing tension between locally calibrated measurements of $H_0$ and its value inferred within the standard $\Lambda$CDM framework. The corresponding estimation of $\chi^{2}_{SN}$ is expressed as \cite{Conley_2010Apjs} 
\begin{equation}\label{chisquare_pantheon}
 \chi^2_{\text{SN}} = \left(\Delta\mu(z_i, \Theta)\right)^{T} C_{SN}^{-1} \left(\Delta\mu(z_i, \Theta)\right)\,,
\end{equation}

where $C_{SN}$ is the relevant covariance matrix that incorporates both systematic and statistical uncertainties, and $ \Delta\mu(z_i, \Theta) = \mu(z_i, \Theta) - \mu(z_i)_{\text{obs}}$.

\textbf{ DESI DR2 BAO data set :} Numerous surveys have greatly improved the understanding of cosmic distance scales via Baryon Acoustic Oscillation (BAO), such as the six-degree field Galaxy Survey, which was conducted at an effective redshift of $z_{eff}$ = 0.106 \cite{Beutler_2011baosixdegree}, the BOSS DR11 quasar Lyman-alpha observations at $z_{eff}$ = 2.4 \cite{du_Mas_des_Bourboux_2017}. The SDSS Main Galaxy Sample at $z_{eff}$ = 0.15 \cite{Ross_2015}. Furthermore, the $H(z)$ measurements and the angular diameter distances obtained from the SDSS-IV eBOSS DR14 quasar survey at effective redshifts of $z_{eff} = \{0.98, 1.23, 1.52, 1.94\}$ \cite{Zhao_2018sdss_IV}, along with the agreed-upon BAO measurements of the Hubble parameter and the associated comoving angular diameter distances from the SDSS-III BOSS DR12 at $z_{eff}= \{0.38, 0.51, 0.61\}$ \cite{Alam_2017sdss_III}. In this article, we exclusively utilized BAO data from the DESI Data Release 2 (DR2) sample, which offers some of the most precise geometric measurements of the late-time Universe. The DESI DR2 sample, which includes low and intermediate redshifts from a variety of galaxy tracers, is described in \cite{Abdul_Karim_2025_Desi}. To analyze the BAO data set, we need to define the Hubble distance \(D_H(z)\), the comoving angular diameter distance \(D_M(z)\), and the volume-average distance \(D_V(z)\)
\begin{eqnarray}\label{BAO_distances}
D_H(z) = \frac{c}{H(z)},\quad D_M(z) = (1 + z)D_A(z), \quad D_V(z) = \left[(1 + z)^2 D_A^2(z) \frac{z}{H(z)}\right]^{1/3},
\end{eqnarray}

Here, the angular diameter distance is related to the luminosity distance through \(D_A(z) = (1+z)^{-2} D_L(z)\). To incorporate the BAO observations into the MCMC analysis, we consider the corresponding set of BAO observables given by \( \mathcal{F}(z_i) = \bigg\{\frac{D_V(z_i)}{r_s(z_d)}, \frac{r_s(z_d)}{D_V(z_i)}, D_H(z_i),D_M(z_i)\bigg(\frac{r_{s,\text{fid}}(z_d)}{r_s(z_d)}\bigg),\\ H(z_i)\bigg(\frac{r_s(z_d)}{r_{s, \text{fid}}(z_d)}\bigg), D_A(z_i)\bigg(\frac{r_{s, \text{fid}}(z_d)}{r_s(z_d)}\bigg)\bigg\} \). To evaluate these observables, we needed to compute the comoving sound horizon \(r_s(z)\) at the redshift \(z_d \approx 1059.94\) \cite{Planck:2018vyg} after the baryon drag epoch.
\begin{eqnarray}\label{sound_horizon}
r_s(z) = \int_{z}^{\infty} \frac{c_s(\tilde{z})}{H(\tilde{z})} d\tilde{z} =\frac{1}{\sqrt{3}} \int_{0}^{1/(1+z)} \frac{da}{a^2 H(a) \sqrt{1 + \left[\frac{3\Omega_{b,0}}{4\Omega_{\gamma,0}}\right] a}},
\end{eqnarray}

In the present analysis, we adopt the values $\Omega_{b,0} = 0.02242$ \cite{Planck:2018vyg}, $T_0 = 2.7255 \, \text{K}$ \cite{Fixsen_2009temcmb} and a fiducial value of $r_{s, \text{fid}}(z_d) = 147.78 \, \text{Mpc}$. The $\chi^{2}$ statistic associated with the BAO observations is defined as \cite{Conley_2010Apjs}
\begin{equation}\label{chisquare_bao}
\chi^2_{\text{BAO}}(\Theta) = \left(\Delta \mathcal{F}(z_i, \Theta)\right)^T C^{-1}_{\text{BAO}} \Delta \mathcal{F}(z_i, \Theta)\,.
\end{equation}

The $C_{\text{BAO}}$ denotes the covariance matrix corresponding to the DESI DR2 BAO observations, and $\Delta \mathcal{F}(z_i, \Theta)$ is defined as $\mathcal{F}(z_i, \Theta) - \mathcal{F}_{\text{obs}}(z_i)$.

 In our analysis, the set of free parameters is given by $(\Theta={H_0,\Omega_{m0},\alpha,\beta,M})$, for which appropriate uniform prior ranges are adopted to explore the allowed parameter space. The priors are imposed as
\begin{equation}
\begin{aligned}
&H_0 \in [50,80],\
&\Omega_{m0}\in[0,0.5],\
&\alpha\in[-5,-1.5],\
&\beta\in[2.5,10],\
&M\in[-50,50].
\end{aligned}
\end{equation}
These prior ranges are chosen sufficiently broad to allow the observational data to constrain the model parameters without imposing unnecessarily restrictive assumptions on their posterior distributions. We sample the posterior distribution using the affine-invariant ensemble sampler implemented in emcee, employing 50 walkers. The walkers are first evolved for 1000 burn-in steps, after which the sampler is reset, and the production chains are evolved for up to $N=100,000$ steps. Convergence is monitored through the integrated autocorrelation time $\tau_{int}$. The chains are considered converged when the total number of production steps exceeds 100 times the estimated autocorrelation time for all parameters, $N > 100 \tau_{int}$, and the relative change in the autocorrelation-time estimate between successive evaluations is smaller than 1\%. The final parameter constraints are obtained from the converged production chains after marginalizing over the remaining model parameters. To assess the sensitivity of the inferred constraints to the adopted prior range, we repeat the analysis with an extended prior on $\alpha$, namely $-8 < \alpha < -1.5$, while keeping the priors on the other parameters unchanged. We find that the marginalized posterior constraints remain nearly unchanged under the two prior choices, indicating that the inferred parameter values are not significantly influenced by the lower boundary of the prior, $\alpha=-5$.

We assess the statistical effectiveness of the cosmological model by determining the minimum $\chi^2_{\rm min}$ value, which is derived from the maximum likelihood using the formula $\chi^2_{\rm min} = -2\ln L_{\rm max}$. To contrast the proposed $f(Q, B)$ gravity model with the conventional $\Lambda$CDM framework, we utilize the Akaike Information Criterion (AIC), which considers both the fit quality and the complexity of the model based on the number of free parameters $k$. The AIC is defined as
\begin{equation}\label{AIC}
\text{AIC} = \chi^{2}_{min} + 2k\,.   
\end{equation}

We also investigate the Bayesian Information Criterion (BIC), which is similar to AIC but places more focus on model complexity than AIC does, and is represented as
\begin{equation}\label{BIC}
\text{BIC} = \chi^{2}_{min} + k \ln N \,.    
\end{equation}

Where $N$ denotes the total number of data points. Like the AIC, the BIC evaluates the trade-off between model fit and complexity, but it applies a more substantial penalty for extra parameters due to its reliance on $\ln N$.

To evaluate the performance of models using various combinations of data sets, we compute the differences in AIC and BIC relative to the reference model, $\Lambda$CDM. The constrained parameters for the $\Lambda$CDM model for each data set combination are detailed in {\color{blue}Appendix}. Lower values of $\Delta$AIC and $\Delta$BIC indicate that the selected model fits the data set more closely to the $\Lambda$CDM model, signifying improved performance. The definitions for $\Delta$AIC and $\Delta$BIC are as follows,
\begin{align}
\Delta\text{AIC} &= \Delta \chi^{2}_{min} + 2\,\Delta k\,,\\
\Delta\text{BIC} &= \Delta \chi^{2}_{min} + \Delta k\, \ln N\,.
\end{align}

%%%%%%%%%%%%%%%%%%%%%%%%%%%%%%%%%%%%%%%%%%%
%%%%%%%%%%%
\subsubsection{MCMC Parameter Estimation and Posterior Distributions}

\begin{figure}[htb]
 \centering
 \includegraphics[width=85mm]{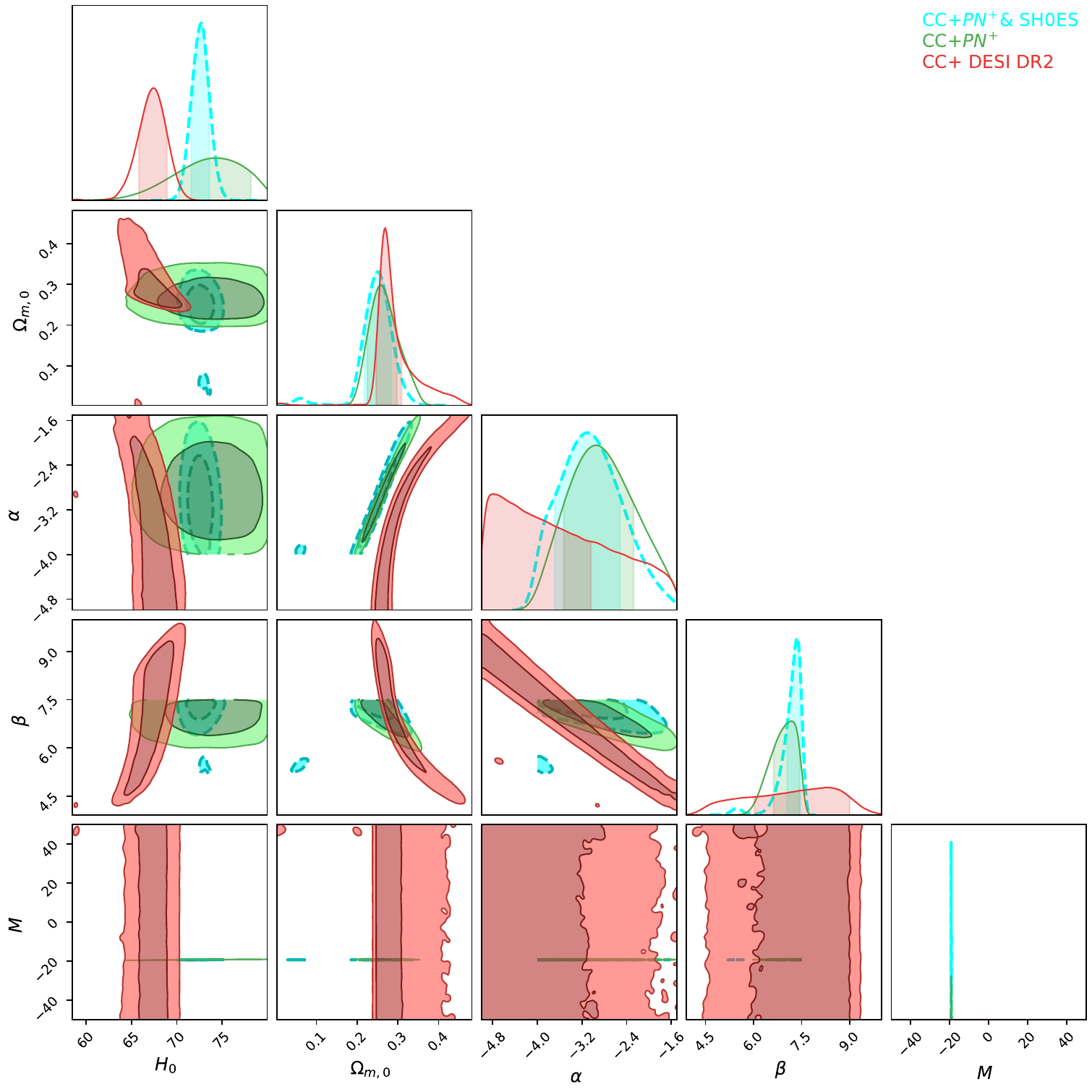}
 \caption{The marginalized posterior distributions and the corresponding 68\% and 95\% confidence contours for the $f(Q,B)$ model parameters, derived from the different dataset combinations.} \label{plot:modelmcmc}
 \end{figure}

In this section, we have established confidence regions and posterior distributions for the model parameters across various combinations of observational datasets. Fig.~\ref{plot:modelmcmc} illustrates the marginalized one-dimensional posterior distributions and two-dimensional confidence contours derived from the MCMC analysis for the different dataset combinations, which include the PN$^{+}$, PN$^{+}$\& SH0ES, and DESI DR2 datasets. The inner and outer contours represent the $68\%$ and $95\%$ confidence levels, respectively. These contour plots provide a comprehensive assessment of the statistical uncertainties and parameter correlations associated with the model. The corresponding parameter constraints obtained from different combinations of observational datasets are summarized in Table~\ref{model_outputs}. The constraints indicate that the inclusion of different observational probes significantly affects the allowed parameter regions, particularly for the Hubble constant $H_0$ and the model parameters $\alpha$ and $\beta$.

For CC+$PN^{+}$ datasets, we find that $H_0=74.4^{+3.8}_{-4.1}$ km s$^{-1}$ Mpc$^{-1}$, indicating a relatively high value for the Hubble constant. Nevertheless, the uncertainty is still quite substantial due to the limited constraining ability of this specific dataset combination. The incorporation of the SH0ES measurement greatly enhances the constraints on cosmological parameters. For CC+$PN^{+}$\&SH0ES combination, the Hubble constant is measured at $H_0=72.70^{+0.91}_{-1.09}$ km s$^{-1}$ Mpc$^{-1}$, reflecting a significant decrease in uncertainty compared to the CC+$PN^{+}$ scenario. This highlights the robust constraining power of the SH0ES measurement regarding the current rate of expansion. This observation aligns with the elevated $H_0$ value reported by the SH0ES distance-ladder measurement, which states $H_0 = 73.04 \pm 1.04 \,\text{km s}^{-1} \, \text{Mpc}^{-1}$ \cite{Riess_2022panplus}. The matter density parameter aligns well with findings from the CC+$PN^{+}$ combination, resulting in $\Omega_{m0}=0.251^{+0.032}_{-0.033}$. The parameters $\alpha$ and $\beta$ are estimated as $\alpha=-3.11^{+0.59}_{-0.58}$ and $\beta=7.36^{+0.11}_{-0.31}$, respectively, while the nuisance parameter is closely constrained around $M=-19.244^{+0.029}_{-0.030}$. The combination of the CC+DESI DR2 dataset yields a lower estimate for the Hubble constant, $H_0=67.5^{+1.4}_{-1.6}$ km s$^{-1}$ Mpc$^{-1}$, which aligns with the value derived from early-Universe constraints within the standard $\Lambda$CDM model. As usual,  the nuisance parameter $M$ remains unconstrained for this combination. The estimated values of the Hubble constant $H_0$ are align with the Planck 2018 $\Lambda$CDM constraint, $H_0=67.4\pm0.5$ km s$^{-1}$ Mpc$^{-1}$ \cite{Planck:2018vyg}. In Table~\ref{model_outputAICBIC}, we display the statistical findings, which include $\chi^{2}_{\text{min}}$, $\Delta$AIC, and $\Delta$BIC for various combinations of data sets. From the obtained $\Delta$AIC and $\Delta$BIC values, we find that the CC+DESI DR2 dataset combination yields the smallest differences compared with the $\Lambda$CDM model among all considered datasets. This indicates that, for this observational combination, the $f(Q, B)$ gravity model achieves statistical performance comparable to the standard $\Lambda$CDM scenario while remaining consistent with the observational data. We further note that the reduced chi-squared for the CC+DESI DR2 combination is $\frac{\chi^2_{\min}}{\mathrm{dof}}=\frac{16.49}{43-5}\simeq 0.43$, where $\mathrm{dof}=N_{\mathrm{data\, sample}}-N_{\mathrm{model\, parameter}}$ denotes the number of degrees of freedom. The corresponding value for the $\Lambda$CDM reference model is $\frac{\chi^2_{\min}}{\mathrm{dof}}=\frac{19.08}{43-3}\simeq 0.48$. Thus, both models yield reduced chi-squared values below unity, indicating that the residuals are small relative to the adopted observational uncertainties. The comparable values obtained for the $f(Q,B)$ gravity and $\Lambda$CDM models further indicate that the low reduced chi-squared is not specific to the additional parameters of the present model.
%%%%%%%%%%%%%%%%%%%%%%%%%%%%%%%%%%%%%%%%%%%%%%%%%%%
%%%%%%%%%%%%%%%%%%%%%
\begin{table}
 \renewcommand{\arraystretch}{1.5}
    \centering
    \caption{The table displays results related to the $f(Q, B)$ model, where the first column identifies the combinations of data sets. The second column indicates the Hubble constant $H_0$, while the third and fourth columns provide the values for the matter density $\Omega_{m0}$ and the model parameter $\alpha$, respectively. The fifth column represents $\beta$. The sixth column presents the Nuisance parameter $M$.}
    \label{model_outputs}
    \begin{tabular}{cccccc}
        \hline
		Data sets & $H_0\, \text{km} \, \text{s}^{-1} \text{Mpc}^{-1}$ & $\Omega_{m0}$ & $\alpha$ & $\beta$ &$M$ \\ 
		\hline
        CC+$PN^{+}$ & $74.4^{+3.8}_{-4.1}$ & $0.258^{+0.040}_{-0.033}$ & $-2.95^{+0.68}_{-0.58}$ & $7.18^{+0.25}_{-0.56}$ & $-19.17^{+0.11}_{-0.13}$ \\
        CC+$PN^{+}\&$ SH0ES & $72.70^{+0.91}_{-1.09}$ & $0.251^{+0.032}_{-0.033}$ & $-3.11^{+0.59}_{-0.58}$ & $7.36^{+0.11}_{-0.31}$ & $-19.244^{+0.029}_{-0.030}$ \\ 
		CC+DESI DR2 & $67.5^{+1.4}_{-1.6}$ & $0.269^{+0.040}_{-0.022}$ & $-4.956^{+1.920}_{-0.033}$ & $8.28^{+0.72}_{-2.25}$ & Unconstrained \\ 
		\hline
    \end{tabular}
\end{table}
%%%%%%%%%%%%%%%%%%
\begin{table}[h]
\renewcommand{\arraystretch}{2}
    \centering
    \caption{This table provides a statistical comparison between the chosen model and the standard $\Lambda$CDM model. More details about the $\Lambda$CDM model can be found in {\color{blue}Appendix}.  The first column displays the data sets. The second column displays the samples size (N) of the data set. For the CC, PN$^{+}$, and DESI DR2 data sets, the corresponding sample sizes are $N=31$, $N=1701$, and $N=12$, respectively. The third column presents the values of $\chi^{2}_{\text{min}}$. The fourth and fifth columns represent the values for $\Delta \text{AIC}$ and $\Delta \text{BIC}$.}
    \label{model_outputAICBIC}
      \begin{tabular}{ccccc}
        \hline
		Data set& N&$\chi^{2}_{min}$ &$\Delta$AIC &$\Delta$BIC \\ 
		\hline
        CC+$PN^{+}$&1732&1800.45 & 12.15&23.34 \\ 	
        CC+$PN^{+}\& SH0ES$&1732&1552.46 &17.24 & 28.15\\ 	
		CC+DESI DR2& 43&16.49 &1.41 & 4.93 \\
  \hline
    \end{tabular}
\end{table}
%%%%%%%%%%%%%%%%%%%%%%%%%%%%%%%%%%%
%%%%%%%%%%%%%%%%%%%%%%%%%%%%%%%%%%
\subsubsection{Parameter Degeneracies: Covariance and Correlation Analysis}

After completing the MCMC analysis, we construct the posterior covariance and correlation matrices to investigate the statistical relationships and parameter degeneracies among the model parameters. For two parameters, $\theta_i$ and $\theta_j$, the posterior covariance is defined as
\begin{equation}
C_{ij}=\frac{1}{N-1}\sum_{k=1}^{N}
(\theta_i^{k}-\bar{\theta}_i)
(\theta_j^{k}-\bar{\theta}_j)\,,
\end{equation}

where $N$ denotes the number of retained MCMC samples, $\theta_i^{k}$ represents the $k$-th realization of the $i$-th parameter, and $\bar{\theta}_i$ is its corresponding posterior mean. The diagonal elements, $C_{ij}$ ($i= j$), represent the posterior variances of the individual parameters and therefore quantify their uncertainties, while the off-diagonal elements, $C_{ij}$ ($i\neq j$), describe the joint variation between parameter pairs. A positive covariance indicates that two parameters tend to vary in the same direction, whereas a negative covariance indicates an opposite tendency. Since the covariance retains the physical units and normalization of the parameters, its magnitude alone does not provide a direct measure of the relative strength of parameter correlations.

To quantify the strength and direction of the parameter degeneracies independently of the parameter scales, we therefore consider the normalized correlation matrix,
\begin{equation}
R_{ij}=\frac{C_{ij}}{\sqrt{C_{ii}C_{jj}}}\,.
\end{equation}

The correlation coefficient satisfies $-1\leq R_{ij}\leq1$, with $R_{ij}=1$ and $R_{ij}=-1$ corresponding to perfect positive and negative linear correlations, respectively, while $R_{ij}\simeq0$ indicates negligible linear correlation. Thus, the correlation matrix provides a scale-independent measure of the degree of linear association and parameter degeneracy within the posterior distribution.

\begin{figure}[htb]
 \centering
 \includegraphics[width=50mm]{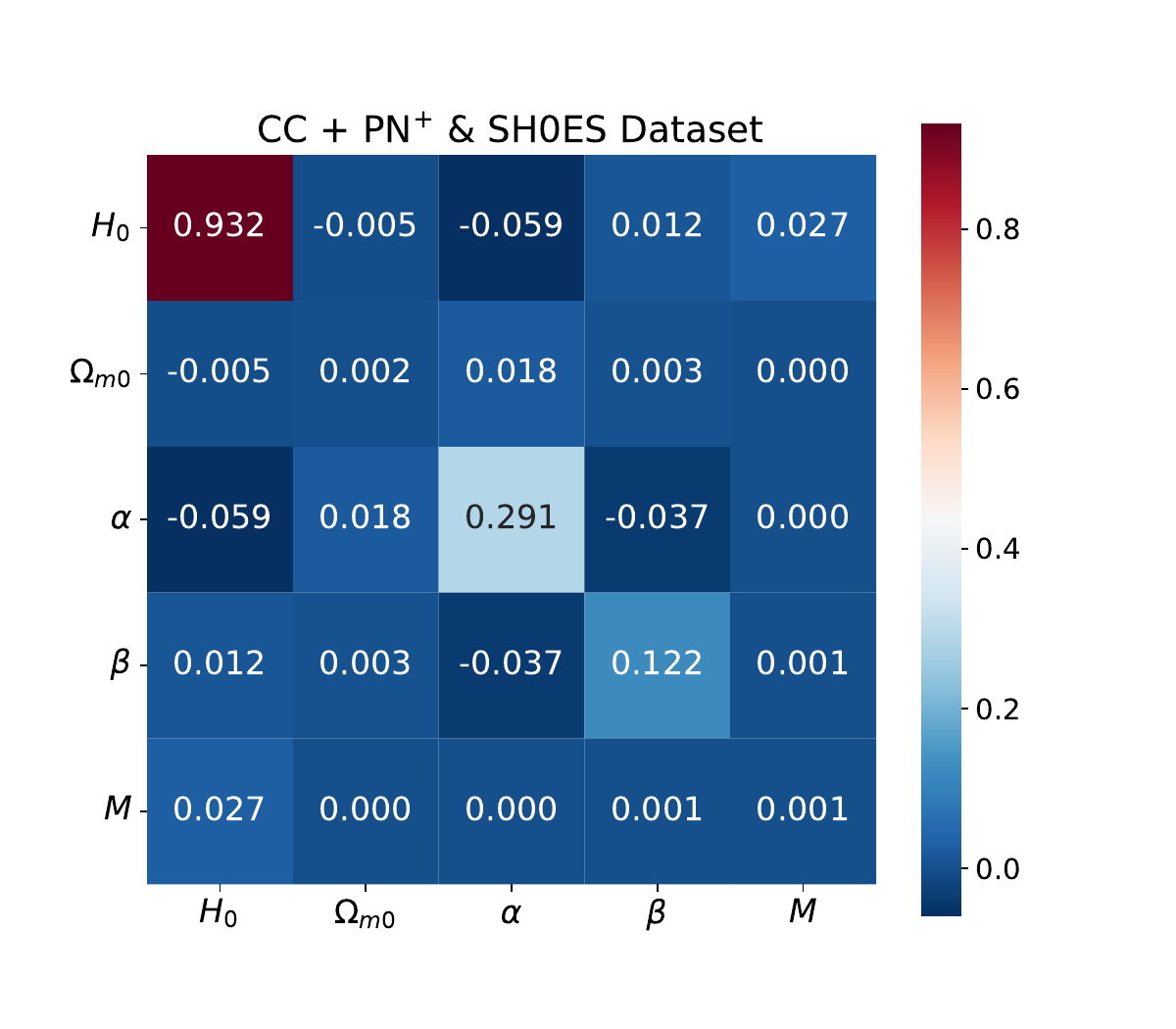}
 \includegraphics[width=50mm]{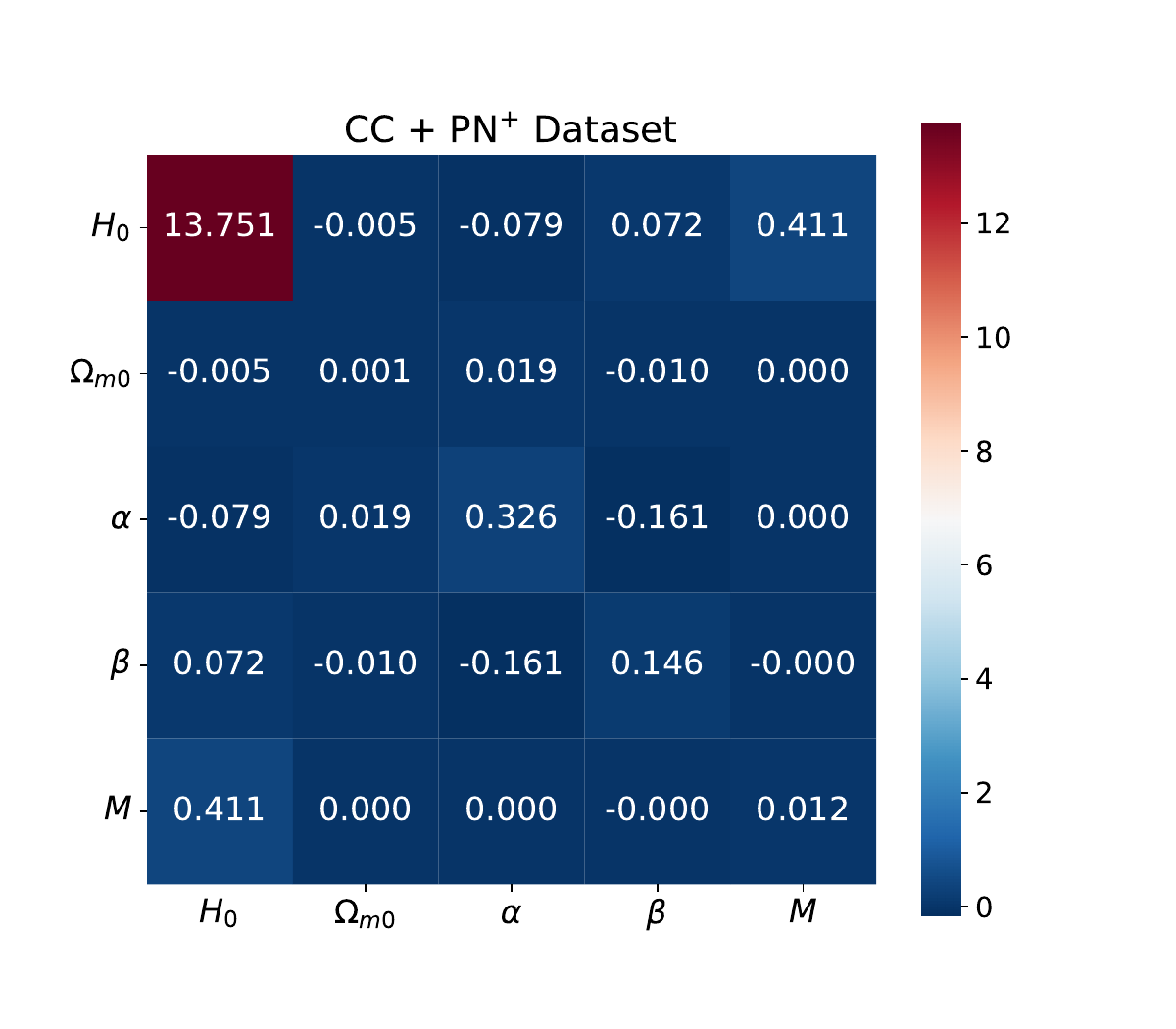}
 \includegraphics[width=50mm]{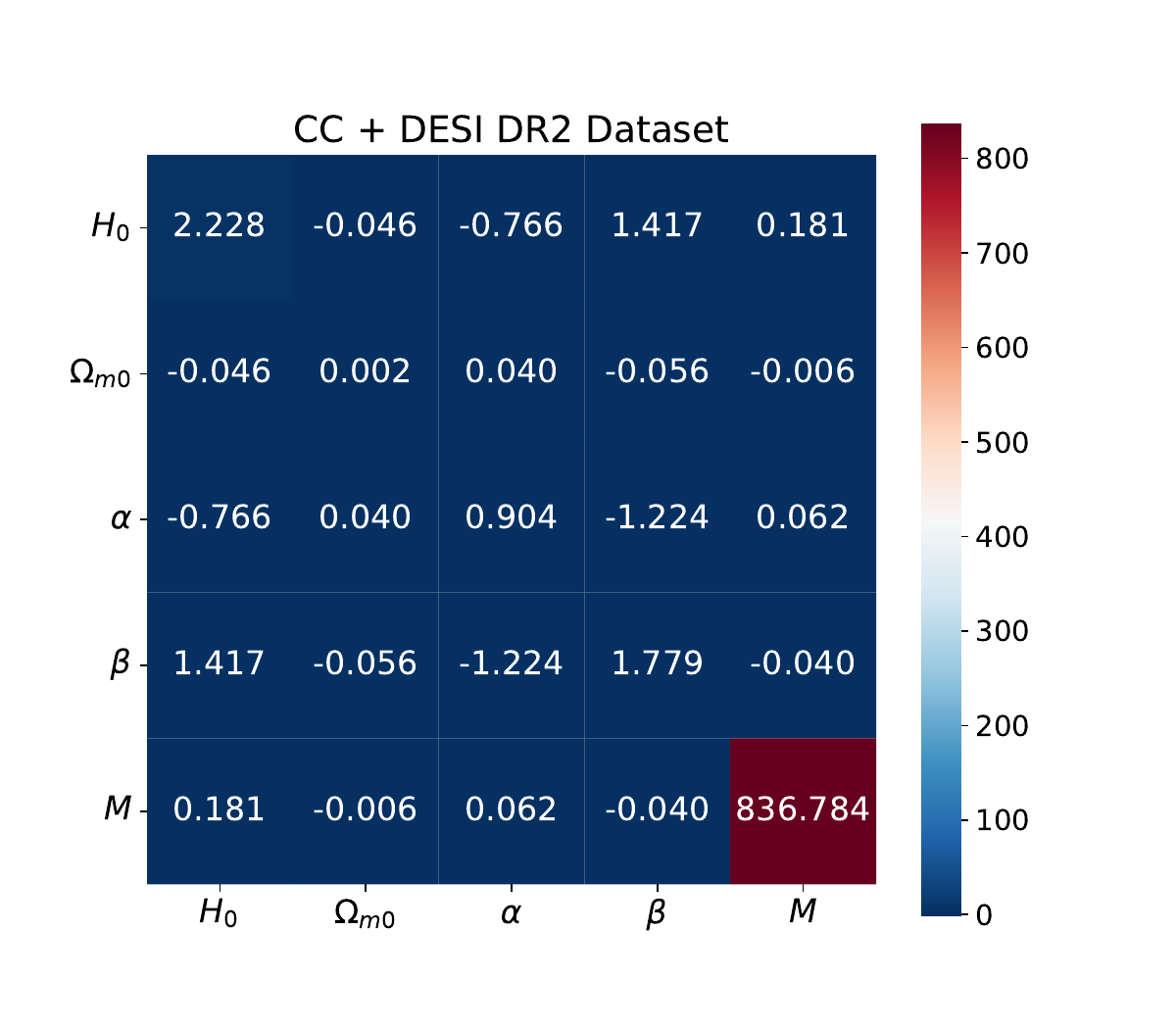}
 \caption{Posterior \textbf{covariance matrices} obtained from the converged MCMC chains for the different combinations of observational datasets. The diagonal elements represent the posterior variances of the parameters, while the off-diagonal elements quantify their joint covariance.} 
 \label{plot:covariancematrix}
 \end{figure}

\begin{figure}[htb]
 \centering
 \includegraphics[width=50mm]{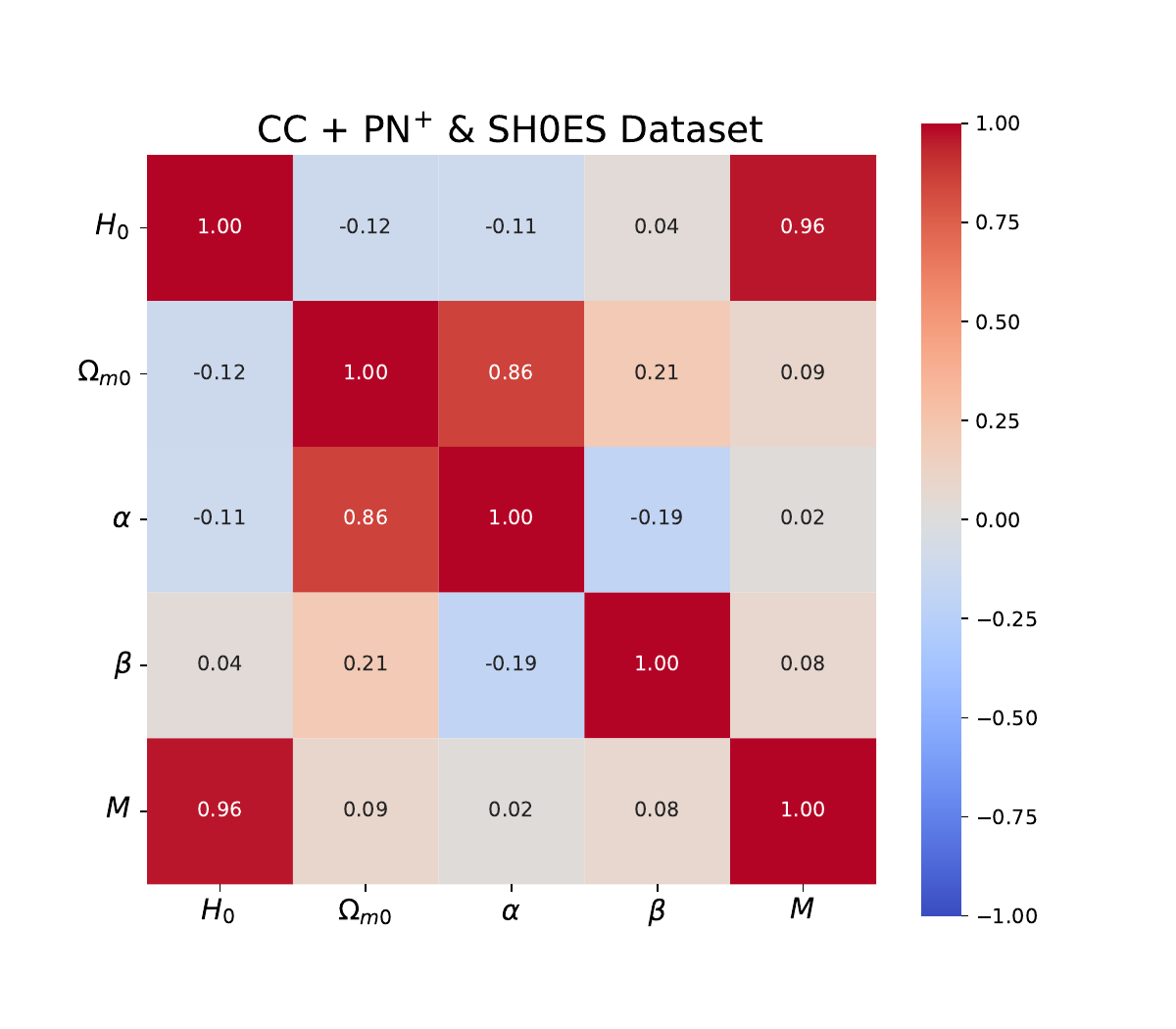}
 \includegraphics[width=50mm]{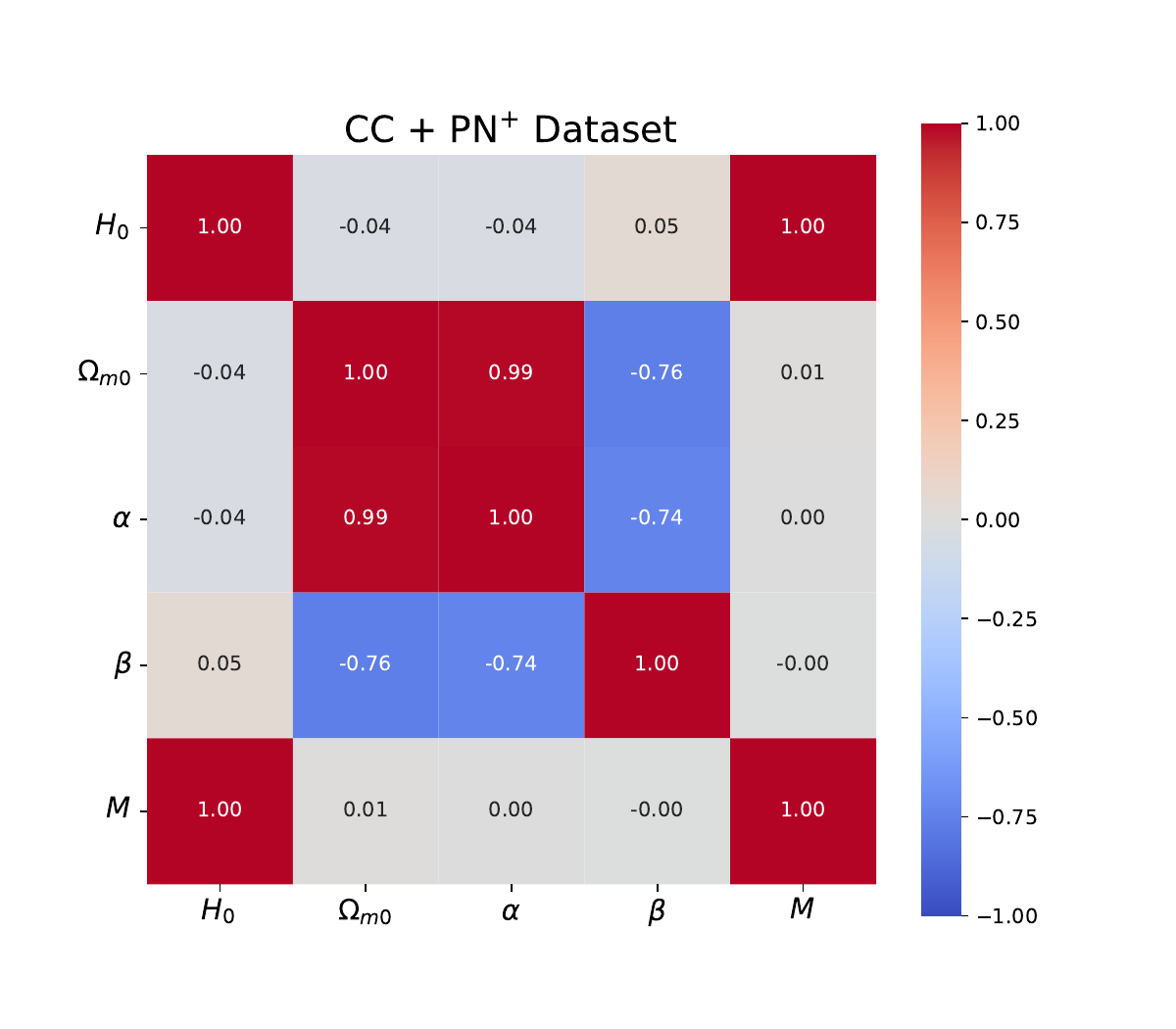}
 \includegraphics[width=50mm]{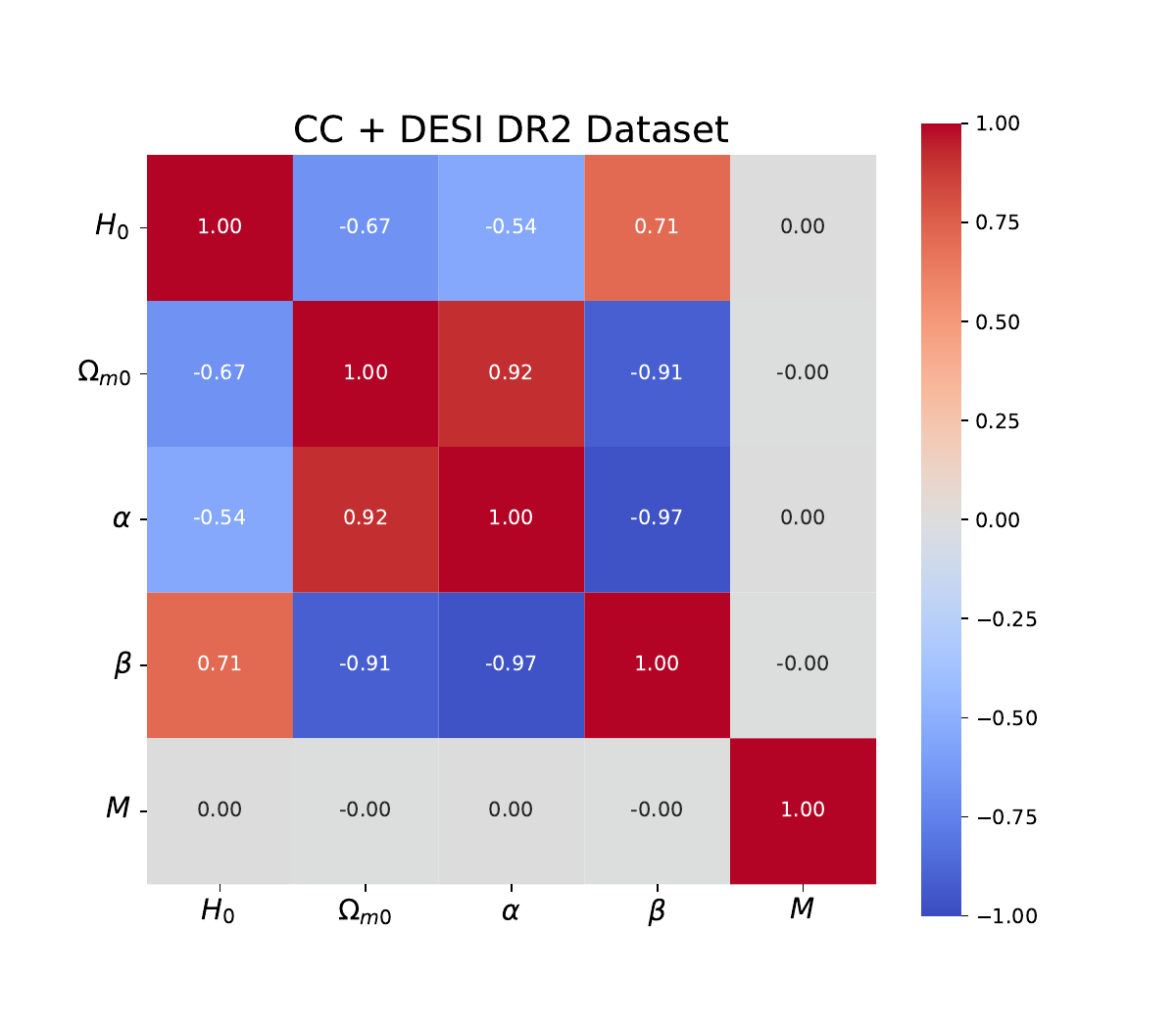}
 \caption{Posterior \textbf{correlation matrices} obtained from the converged MCMC chains for the different combinations of observational datasets. The matrix elements quantify the strength and direction of the posterior correlations between the model parameters, with $R_{ij}=1$ indicating perfect positive correlation, $R_{ij}=-1$ indicating perfect negative correlation, $R_{ij}=0$ no correlation, $R_{ij}>0$ positive correlation and $R_{ij}<0$ negative correlation.}  
 \label{plot:correlationmatrix}
 \end{figure}

In Figs.~\ref{plot:covariancematrix} and \ref{plot:correlationmatrix}, we display the posterior covariance and correlation matrices obtained from the converged MCMC chains for the different combinations of observational datasets. The covariance matrices exhibit an overall diagonal-dominant structure with non-zero off-diagonal elements, indicating that the model parameters are constrained at different levels and possess varying degrees of mutual dependence. The corresponding correlation matrices further reveal the strength and direction of these parameter interdependencies, providing a clearer representation of the underlying posterior degeneracies. In Fig.~\ref{plot:covariancematrix}, positive entries in the covariance matrix suggest that the two parameters move in the same direction, while negative entries imply that they move in opposite directions. Larger values within the covariance matrix signify significant parameter degeneracies, whereas smaller values indicate less pronounced parameter degeneracies. In Fig.~\ref{plot:correlationmatrix}, $R_{ij}=1$ indicates a strong positive correlation between the parameters, whereas $R_{ij}=-1$ indicates a strong negative correlation. A value of $R_{ij}=0$ signifies the absence of any linear correlation between the corresponding parameters. Positive values of the correlation matrix $(R_{ij}>0)$ represent positive correlations between the parameters, while negative values $(R_{ij}<0)$ represent negative correlations between the parameters.

Since the CC+DESI DR2 combination contains no SNe Ia data, the absolute-magnitude nuisance parameter \(M\) is not relevant to this likelihood and is therefore unconstrained. This behavior is reflected in the covariance matrix by the substantially larger diagonal element associated with $M$, indicating a large posterior variance and consequently weak constraint on this parameter. Furthermore, the corresponding correlation coefficients between $M$ and the other model parameters are zero, $R_{ij}=0$, indicating that $M$ has no linear correlation with the remaining parameters for this dataset combination. Therefore, the CC+DESI DR2 observations do not provide sufficient constraining power for $M$, and its posterior uncertainty remains large.

Overall, the covariance and correlation matrices provide a complementary assessment of the parameter constraints obtained from the MCMC analysis. The observed covariance structure and parameter correlations demonstrate the presence of non-trivial degeneracies within the posterior parameter space, while also illustrating the constraining capability of the combined observational datasets. These results provide an additional consistency check on the MCMC parameter estimation and contribute to the assessment of the robustness of the cosmological constraints within the considered $f(Q,B)$ framework.
%%%%%%%%%%%%%%%%%%%%%%%%%%%%%%%%%%%%%%%
%%%%%%%%%%%%%
%%%%%%%%%%%%%%%%%%%%%
%%%%%%%%%%%%%%%%%%%%%%%%%%%%%%
%%%%%%%%%%%%%%%%%%%%%%%%%%%%
\subsubsection{Late-Time Cosmology}

\begin{figure}[htb]
     \centering
         \includegraphics[width=50mm]{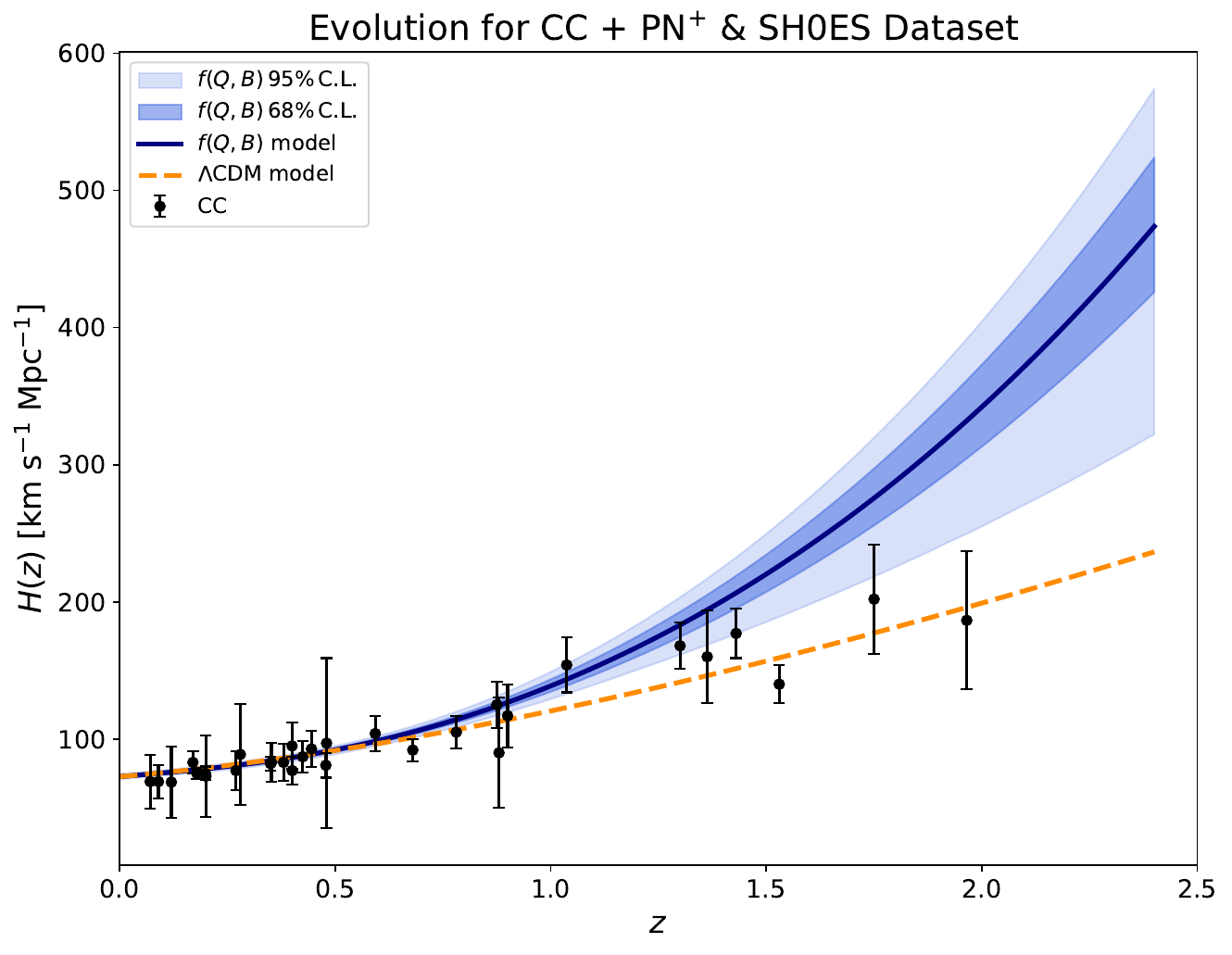}
         \includegraphics[width=50mm]{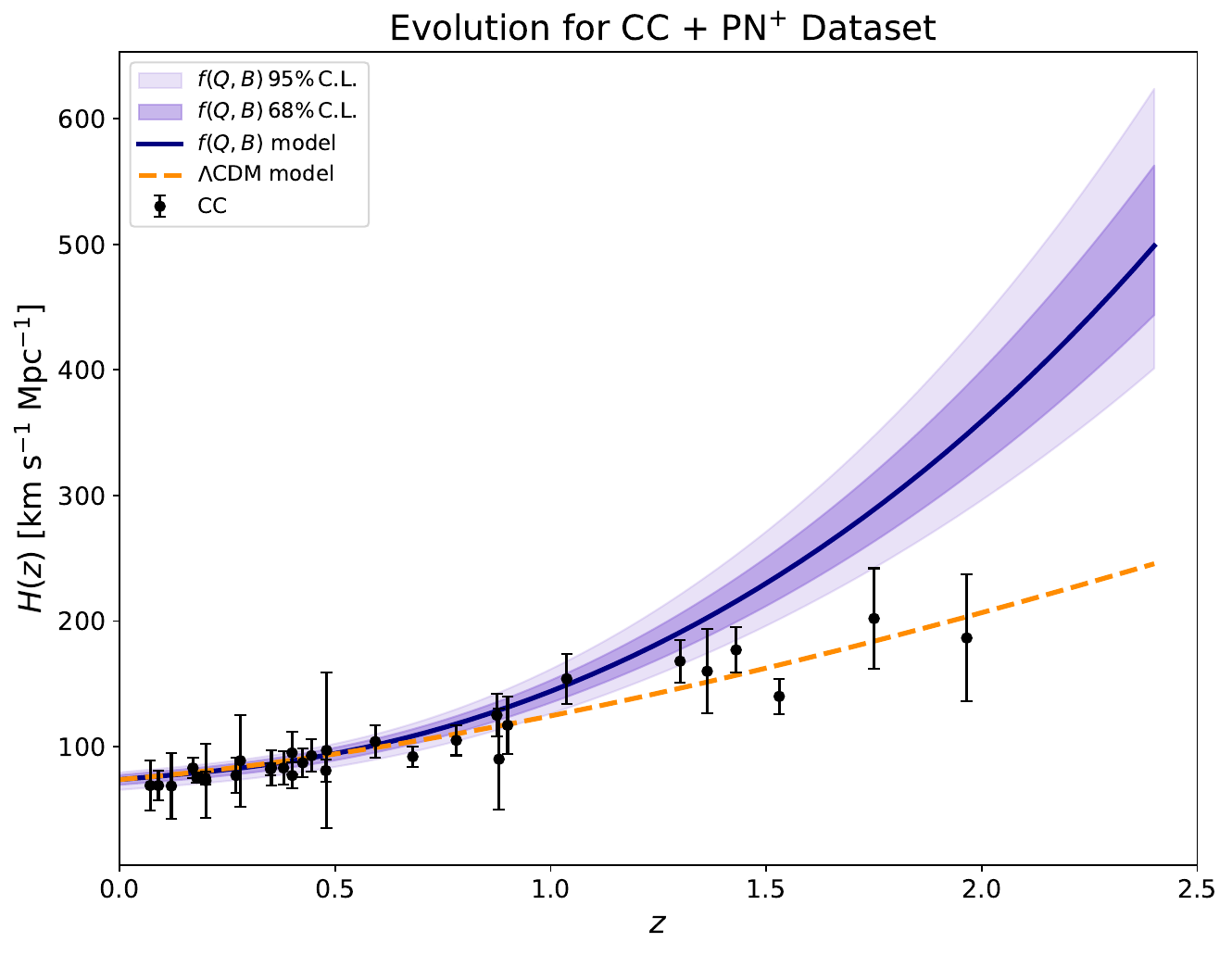}
      \includegraphics[width=50mm]{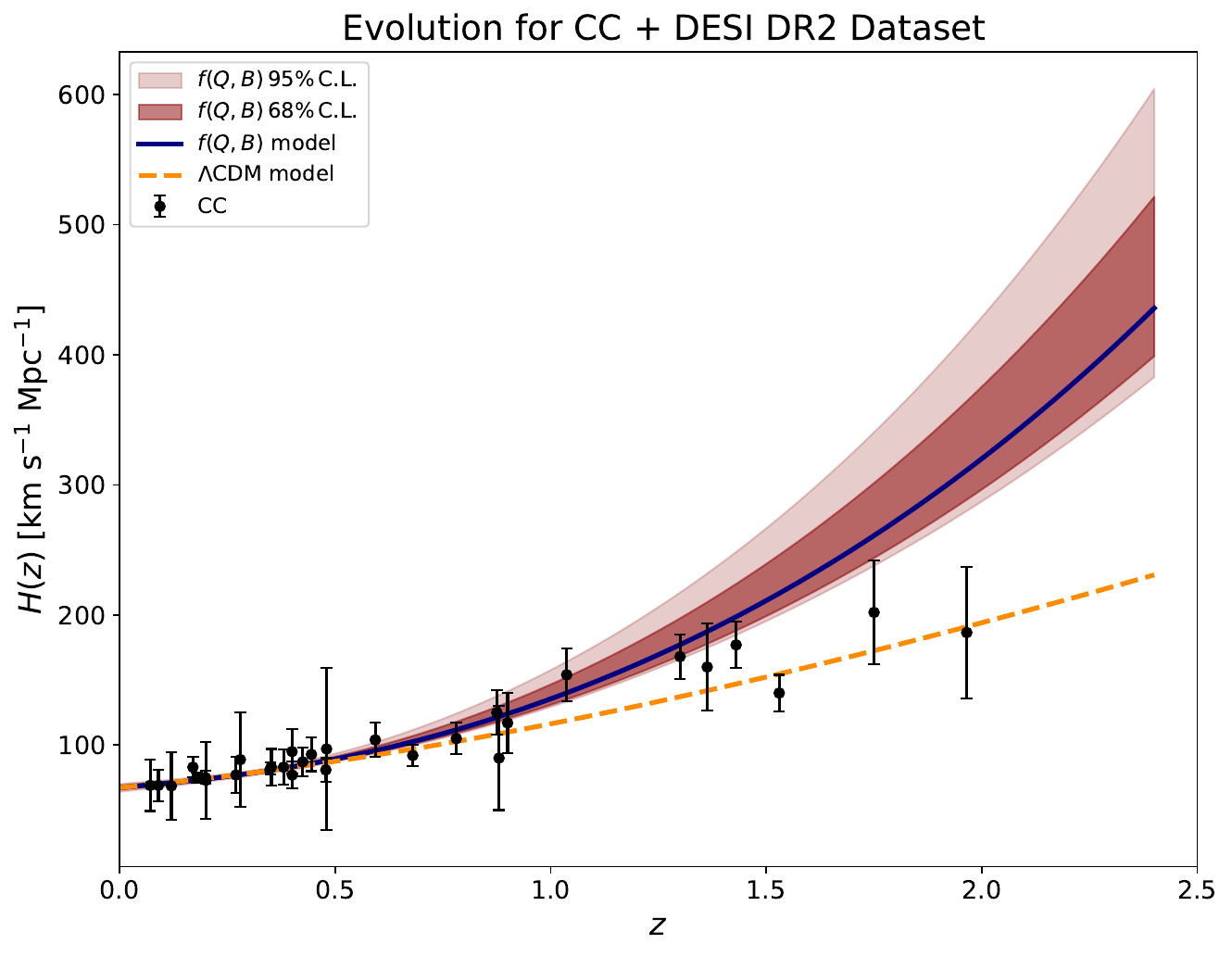}
\caption{Evolutionary behavior of the Hubble parameters and  $\Lambda$CDM model in redshift for the data sets combination.}  
\label{plot:Hubbleplot}
\end{figure}

In this section, we will analyze the fundamental cosmological parameters to explore the late-time dynamics of the $f(Q, B)$ model. In Fig.~\ref{plot:Hubbleplot}, we have shown the behavior of the Hubble parameter for different data set combinations along with the standard $\Lambda$CDM model. For this figure, we used the MCMC-converged chain obtained after running the code. From Fig.~\ref{plot:Hubbleplot}, we can say that for very low redshift, the $f(Q, B)$ model is similar to the standard $\Lambda$CDM model, but as redshift increases, we can see the difference between the model and the $\Lambda$CDM model. However, for the CC+DESI DR2 data set combination, the difference is smaller than for the other two data set combinations. These differences between the $f(Q, B)$ model and the standard $\Lambda$CDM model originate from the inclusion of the nonmetricity scalar ($Q$) and the boundary term ($B$) in the gravitational action. As a result, the cosmological dynamics are modified, leading to a different evolution of the Hubble parameter than in the standard $\Lambda$CDM model. Overall, across all three data sets, the present-time value of the Hubble parameter at $z=0$ is consistent with current cosmological observations. 

\begin{figure}[htb]
     \centering
         \includegraphics[width=50mm]{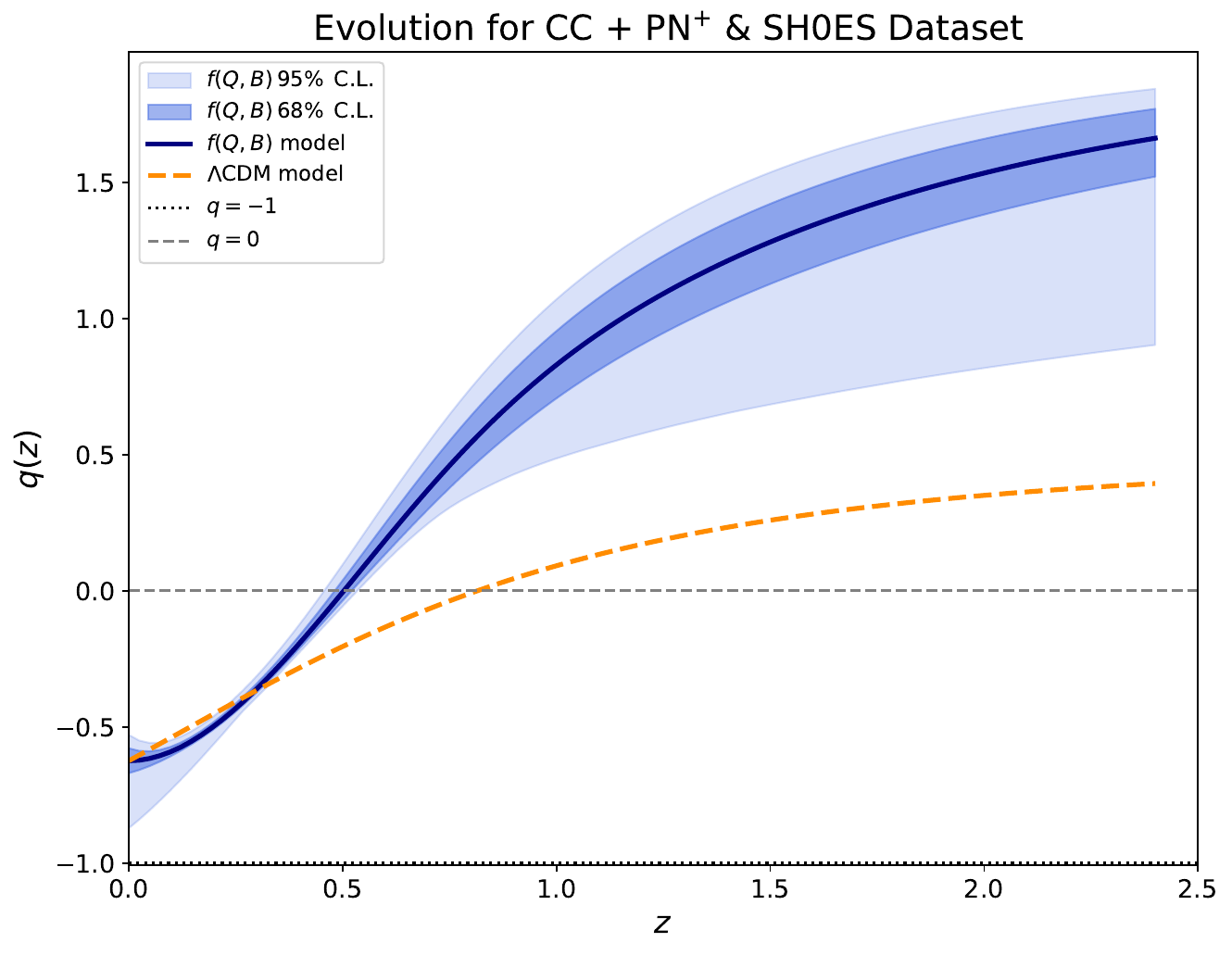}
         \includegraphics[width=50mm]{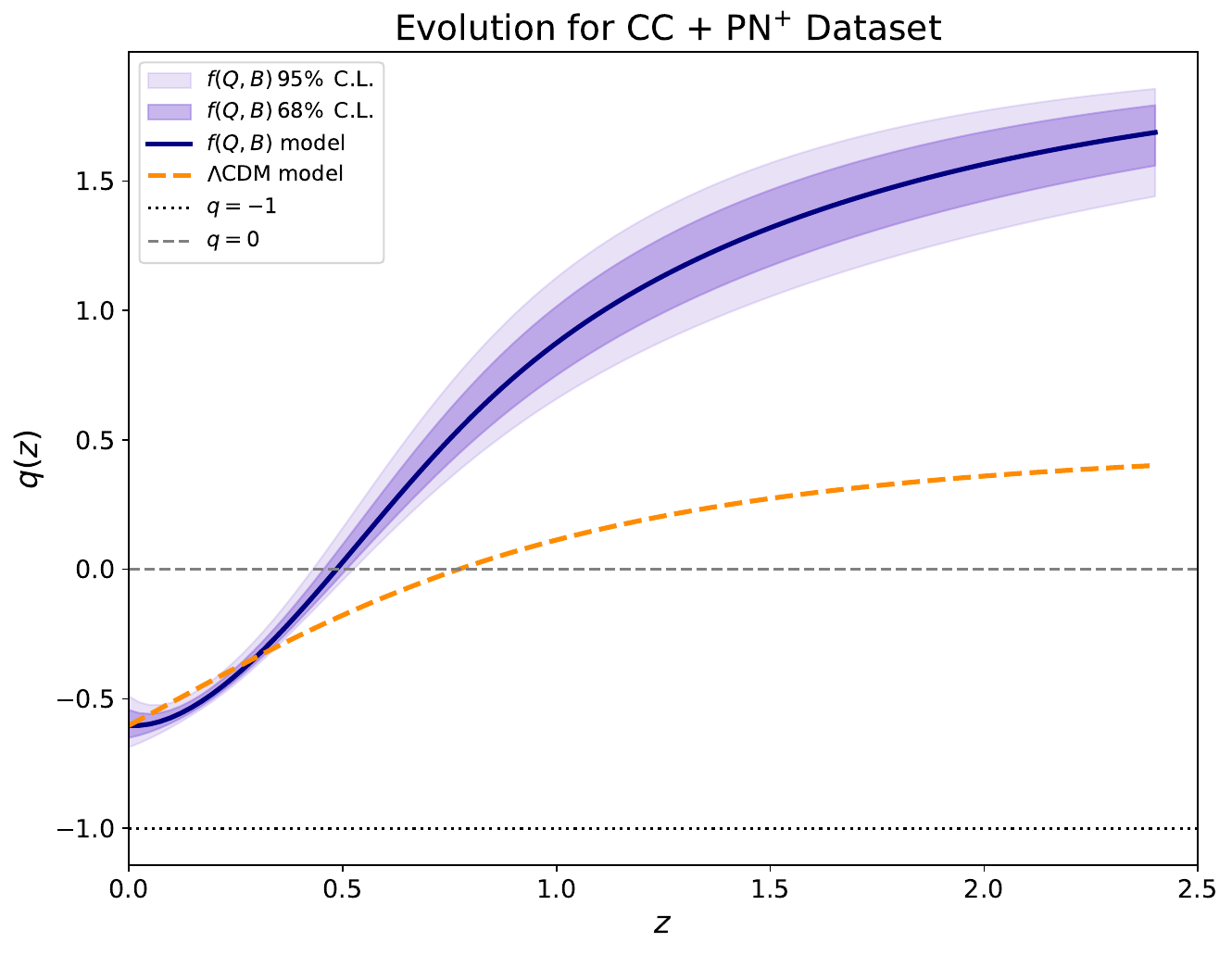}
      \includegraphics[width=50mm]{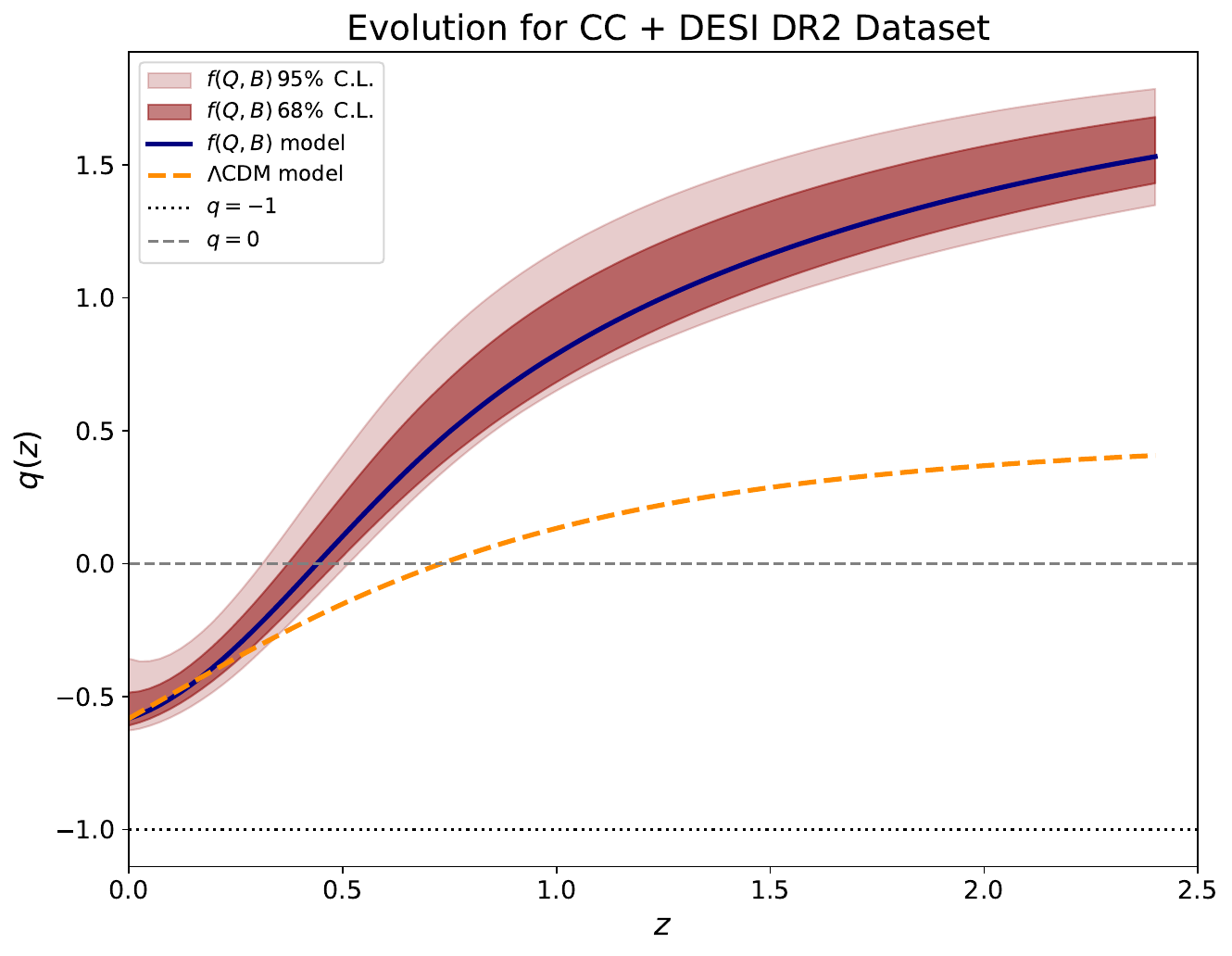}
\caption{Evolutionary behavior of the deceleration parameters and  $\Lambda$CDM model in redshift for the data sets combination. }  
\label{plot:qzplot}
\end{figure}

In Fig.~\ref{plot:qzplot}, we illustrate the evolution in the deceleration parameter $q(z)$ derived from various combinations of observational datasets, alongside the standard $\Lambda$CDM model. In this figure, The negative values of $q(z)$ suggest that the Universe is experiencing accelerated expansion phase, aligning with the observed late-time cosmic acceleration. Based on the behavior of the deceleration parameter depicted in this figure, we can conclude that the \( f(Q, B) \) model indicates the onset of the Universe's acceleration phase occurs slightly later than in the standard \( \Lambda \)CDM model. By analyzing various dataset combinations, we have identified the transition point in the deceleration parameter. This transition point marks the moment when the deceleration parameter shifts from positive to negative, signifying the Universe's entry into an accelerated phase. For the CC + \( PN^{+} \)\&SH0ES dataset, the transition point is found to be \( z_{\text{trans}} = 0.50250^{+0.01809}_{-0.02121} \) (68\% C.L.) and \( z_{\text{trans}}= 0.50250^{+0.03166}_{-0.04594} \) (95\% C.L.). For the CC + \( PN^{+} \) dataset, the corresponding constraints are \( z_{\text{trans}} = 0.48560^{+0.02306}_{-0.03024} \) (68\% C.L.) and \( z_{\text{trans}} = 0.48560^{+0.03936}_{-0.05531} \) (95\% C.L.). Similarly, for the CC + DESI DR2 dataset, we obtain \( z_{\text{trans}} = 0.43979^{+0.04329}_{-0.06827} \) (68\% C.L.) and \( z_{\text{trans}} = 0.43979^{+0.07323}_{-0.12783} \) (95\% C.L.). We also calculate the present-day values of the deceleration parameter, $q_0$, for each set of observational combinations. The present-day value of the deceleration parameter for the CC+$PN^{+}$\&SH0ES dataset is found to be $q_0=-0.62376^{+0.04577}_{-0.04559}\, (68\% $ C.L.) and $q_0=-0.62376^{+0.09514}_{-0.24783}\, (95\% $ C.L.). For the CC+$PN^{+}$ dataset, the corresponding constraints are $q_0=-0.60298^{+0.05982}_{-0.04780}\, (68\% $ C.L.) and $q_0=-0.60298^{+0.11305}_{-0.08425}\, (95\% $ C.L.). Likewise, for the CC+DESI DR2 dataset, we obtain $q_0=-0.58192^{+0.09744}_{-0.02645}\, (68\% $ C.L.) and $q_0=-0.58192^{+0.22524}_{-0.04549}\, (95\% $ C.L.). The results derived from the $f(Q, B)$ model regarding the present value of the deceleration parameter and transition point $z_{\text{trans}}$ align with cosmological observations \cite{Gruber_2014,PhysRevD.90.044016a, PhysRevResearch.2.013028}. The resulting negative present-day values of $q_0$ imply that the Universe is presently experiencing accelerated expansion. These outcomes align with the broader observational understanding of late-time cosmic acceleration and highlight that the investigated $f(Q, B)$ model can replicate the observed accelerated expansion of the current Universe across various combinations of observational datasets.

\begin{figure}[htb]
     \centering
         \includegraphics[width=50mm]{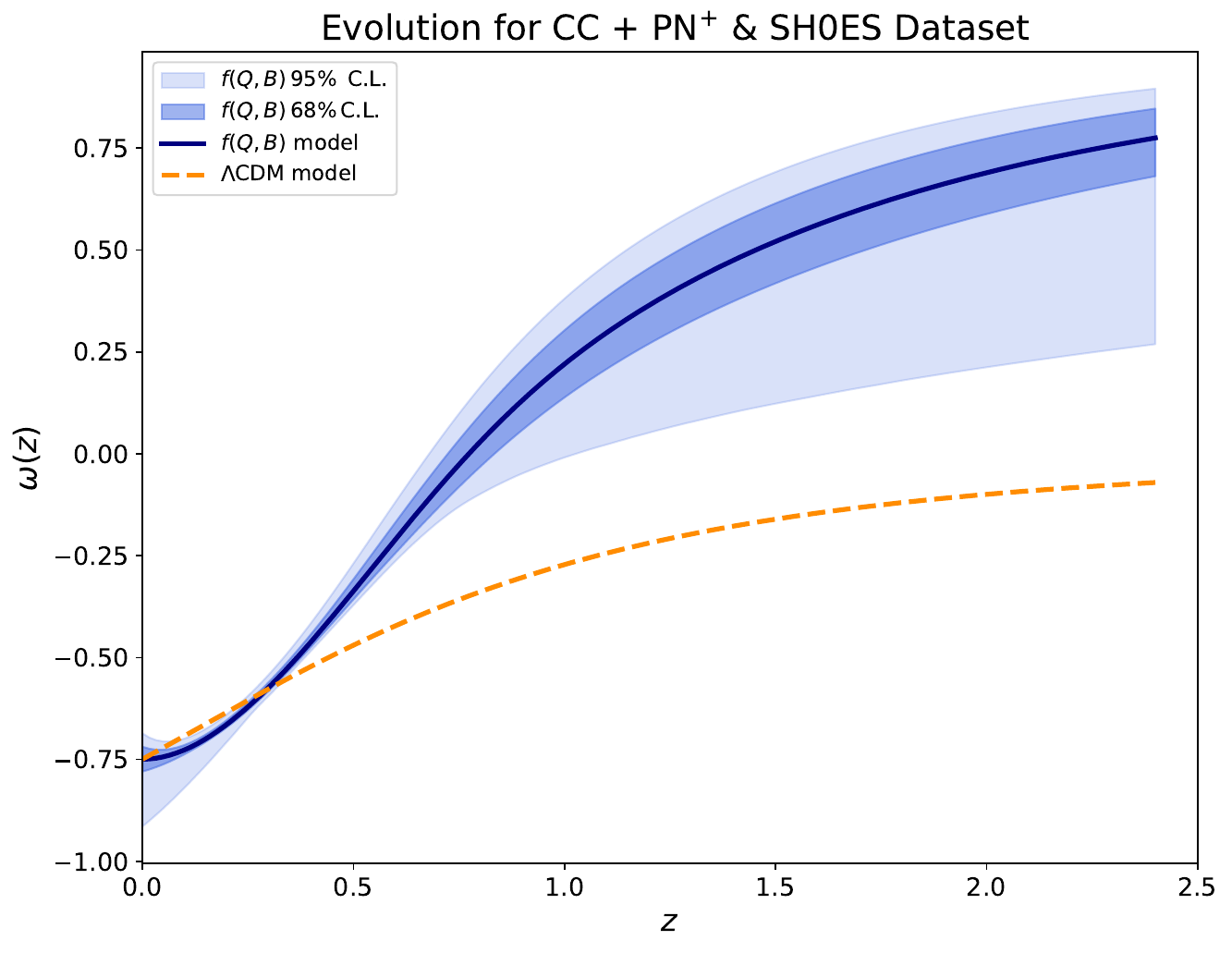}
         \includegraphics[width=50mm]{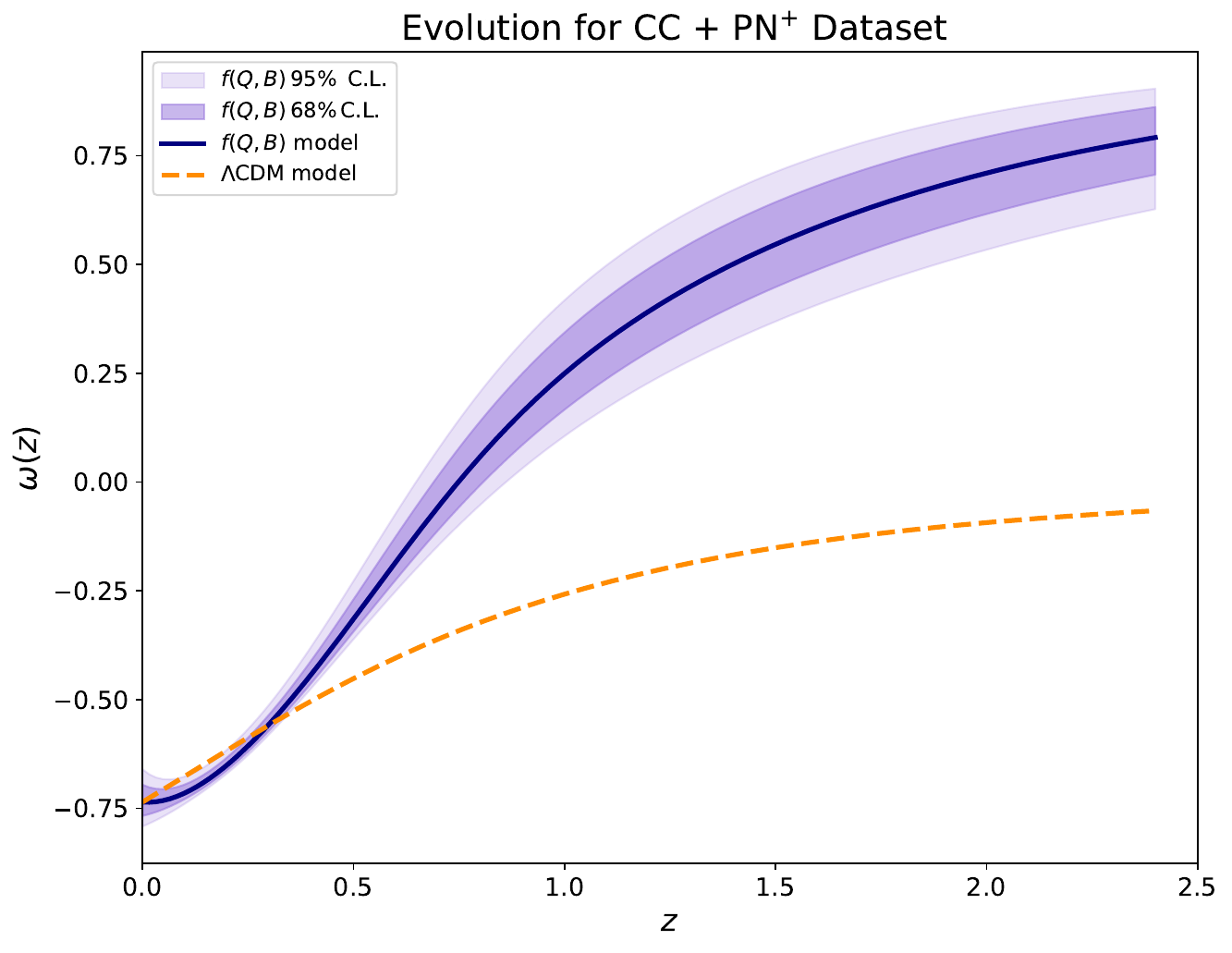}
      \includegraphics[width=50mm]{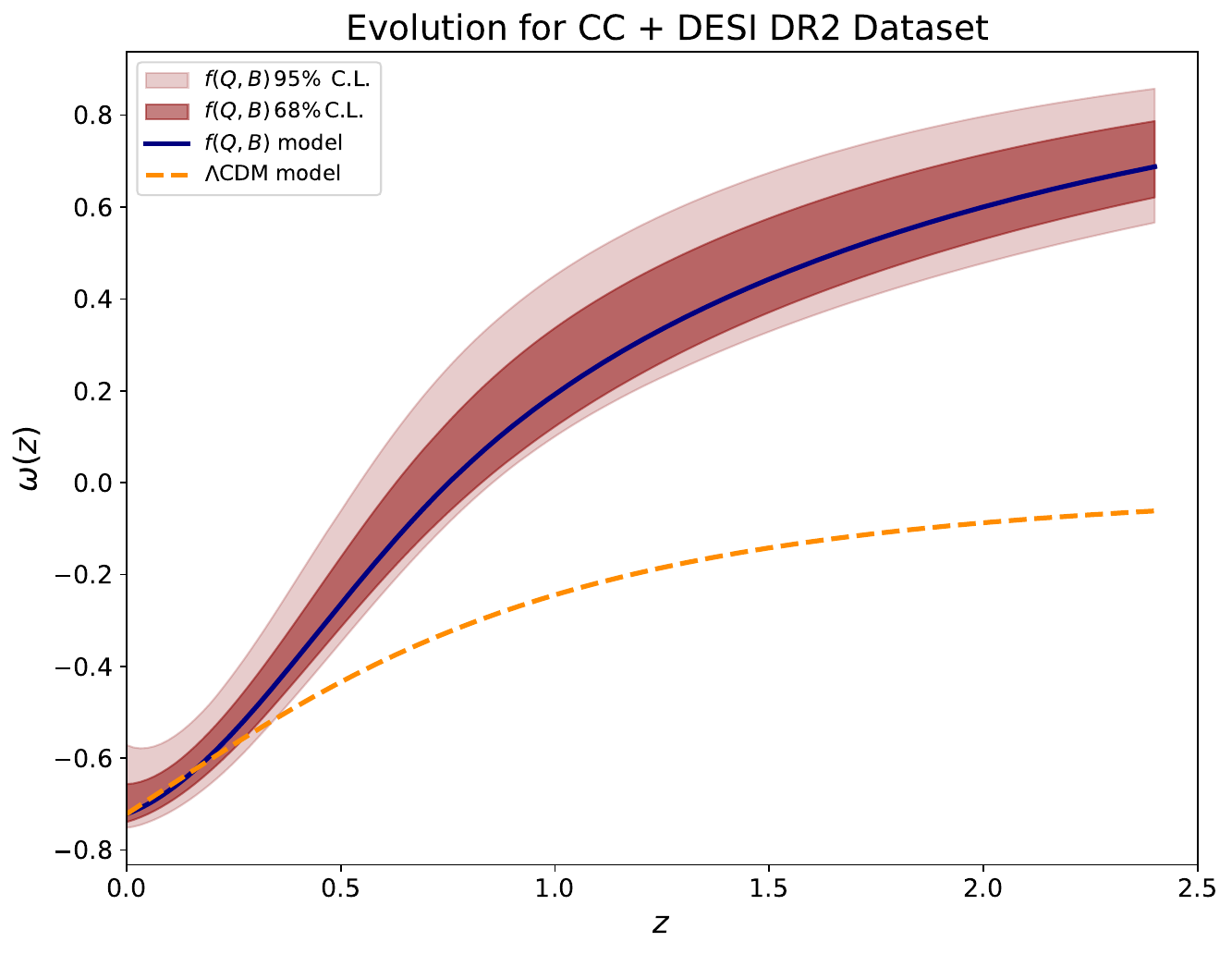}
\caption{Evolutionary behavior of the Equation-of-state parameter and  $\Lambda$CDM model in redshift for the data sets combination. }  
\label{plot:wzplot}
\end{figure}

In Fig.~\ref{plot:wzplot}, we present the evolution of the total equation-of-state (EoS) parameter $\omega(z)$ obtained from the observationally constrained $f(Q, B)$ model for the considered combinations of observational data sets, together with the corresponding $\Lambda$CDM prediction. At low redshifts \( z \lesssim 0.6 \), the EoS remains in the range \( -1 < w(z) < -\frac{1}{3} \), indicating a quintessence-like behavior of the $f(Q, B)$ model and corresponding to accelerated cosmic expansion. After that, $\omega(z)$ approaches $-\frac{1}{3}$ around \( z \simeq 0.6 \) to \( 0.8 \), corresponding to the transition from accelerated to decelerated expansion, and subsequently approaches $\omega(z)\simeq 0$, indicating a matter-like behavior. For \( z \gtrsim 0.8 \), the EoS becomes positive and increases with redshift, with $\omega(z)$ exceeding the radiation-like value $\frac{1}{3}$. Although this behavior corresponds to a strongly positive effective pressure, it should be interpreted with caution. The evolution of $\omega(z)$ is obtained from the numerical solution of the second-order differential equation for $H(z)$, with the boundary conditions specified at $z=0$, and the solution is subsequently evolved toward higher redshifts. The sensitivity of this backward integration to the adopted boundary conditions may therefore influence the behavior of the reconstructed EoS at higher redshifts. Consequently, the high-redshift behavior should be assessed alongside the stability of the numerical solution and its consistency with observational constraints from the data sets used in this analysis. The region in which the reconstructed solution remains physically and numerically reliable should therefore be considered when interpreting the evolution of $\omega(z)$. In particular, the behavior with $\omega(z)>1/3$ should not be interpreted as definitive evidence for a genuine stiff-matter-dominated epoch. Furthermore, the present analysis does not by itself establish the consistency of this high-redshift behavior with the detailed physics of recombination or Big Bang nucleosynthesis. We also determine the current values of the EoS parameter, $\omega_0$, for each set of observational data combinations. The present-day value of the EoS parameter for the CC+$PN^{+}$\&SH0ES dataset is found to be $\omega_0=-0.74918^{+0.03051}_{-0.03039}\, (68\% $ C.L.) and $\omega_0=-0.74918^{+0.06343}_{-0.16522}\, (95\% $ C.L.). For the CC+$PN^{+}$ dataset, the corresponding constraints are $\omega_0=-0.73532^{+0.03988}_{-0.03187}\, (68\% $ C.L.) and $\omega_0=-0.73532^{+0.07537}_{-0.05617}\, (95\% $ C.L.). Likewise, for the CC+DESI DR2 dataset, we obtain $\omega_0=-0.72128^{+0.06496}_{-0.01763}\, (68\% $ C.L.) and $\omega_0=-0.72128^{+0.15016}_{-0.03033}\, (95\% $ C.L.). The results obtained from the $f(Q, B)$ model concerning the current value of the EoS parameter are consistent with cosmological observations \cite{Planck:2018vyg, Hinshaw_2013}.   
%%%%%%%%%%%%%%%%%%%%%%%%%%%%%%%%%%%%%%%%%

%%%%%%%%%%%%%%%%%%%

\begin{figure}[h]
 \centering
 \includegraphics[width=140mm]{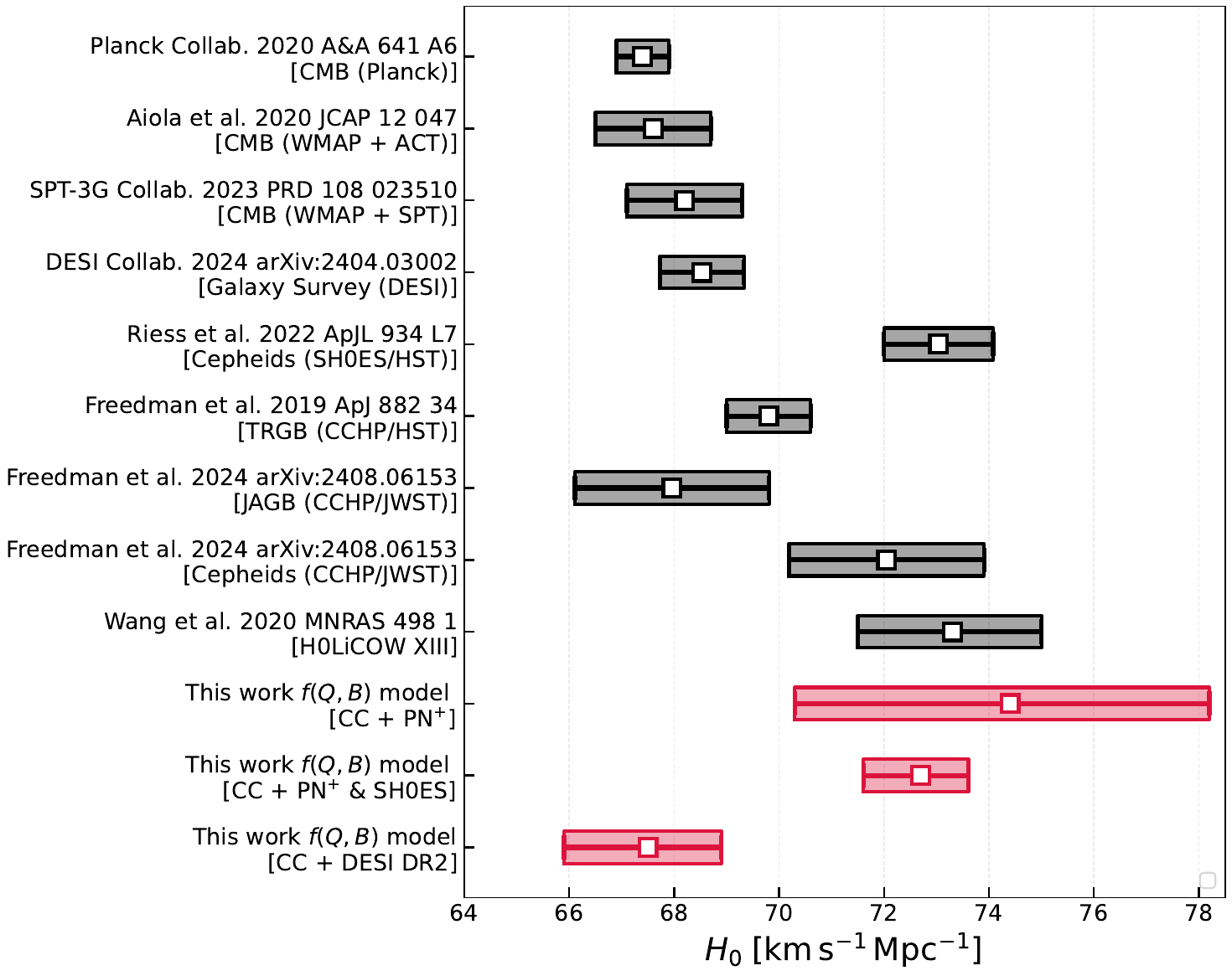} 
 \caption{Whisker plot illustrating a comparison between the observational determinations of the Hubble constant $H_0$ and the constraints obtained from the $f(Q, B)$ gravity model. The black markers represent measurements reported in the literature, whereas the red markers correspond to the results obtained in this work for the CC+PN$^{+}$, CC+PN$^{+}$\& SH0ES, and CC+DESI DR2 dataset combinations. Horizontal error bars denote the $1\sigma$ confidence intervals.
} \label{whiskerplot}
 \end{figure}

 In Fig.~\ref{whiskerplot}, we present a whisker plot comparing the Hubble constant \( H_0 \), derived from various cosmological observations and the latest measurements available in the literature. This includes Cosmic Microwave Background (CMB) observations from Planck, ACT and SPT, results from galaxy surveys conducted by DESI, local distance ladder determinations from SH0ES, TRGB, JAGB and strong lensing studies from H0LiCOW. The plot also showcases the constraints obtained in our current study utilizing the \( f(Q, B) \) gravity model across different datasets, specifically CC+PN\(^+\), CC+PN\(^+\)\& SH0ES and CC+DESI DR2. Each horizontal whisker indicates the corresponding central value alongside its 1$\sigma$ uncertainty, facilitating a direct visual comparison between various \( H_0 \) estimations. The results derived from the \( f(Q, B) \) model occupy an intermediate space between the lower values favored by early-Universe observations, such as those from the CMB, and the higher values inferred from local measurements in the late Universe. Notably, incorporating the SH0ES prior results in an upward shift of the estimated \( H_0 \), enhancing its alignment with local measurements, whereas the CC+DESI DR2 analysis stays closer to the CMB-preferred values. The inferred \(H_0\) is strongly dataset dependent, spanning the range between the early-Universe and locally calibrated determinations.

%%%%%%%%%%%%%%%%%
\section{Conclusion}\label{Sec:conclusion}

In this study, we investigated the Noether symmetry approach and also performed Bayesian cosmological parameter estimation for the power-law $f(Q, B)$ gravity model using recent cosmological observations. The Noether symmetry is one of the most effective mathematical techniques for identifying conserved quantities and simplifying dynamical systems. Employing the Noether symmetry method is advantageous for categorizing models and discovering precise cosmological solutions to the field equations. The Lagrangian is crucial for illustrating symmetries and their corresponding Noether vectors. In this manner, Lagrange multipliers are introduced to impose the definitions of \(Q\) and \(B\) and to construct the point-like Lagrangian, as discussed in Sec ~\ref {SEC:Lagrangian formalism}. We formulate the Lagrangian using the FLRW space-time metric as outlined in Eq.~(\ref{Eq:main_Lagrangian}). To verify the Lagrangian \eqref{Eq:main_Lagrangian}, we have derived the Friedmann equations for the flat FLRW metric from action \eqref{eq: action_fQB}. In Sec~\ref {Sec:NoetherfQB}, we analyze the Rund-Trautman identity presented in Eq.~(\ref{eq:noether_condition}) to derive a system of partial differential equations labeled as Eqs.~(\ref{Eq:system1}--\ref{Eq:system8}), which constitute the governing equations of the system. Within this set of partial differential equations, the unknown variables include $\xi$, $\theta_1$, $\theta_2$, $\theta_3$, and the function $f(Q, B)$. These variables represent Noether coefficients, while $f(Q, B)$ is a general function of the non-metricity and the boundary term. In Sec~\ref{Sec:modelnoether}, we examined the Noether symmetry within the framework of $f(Q, B)$ gravity, focusing on a particular form of $f(Q, B)=f_0 (-Q)^{\alpha}B^{\beta}$. For this specific choice of $f(Q, B)$, we derived a non-trivial Noether vector, which is presented in Eq.~\eqref {eq:powerlaw_scaling_generator}. For this power-law form of the $f(Q, B)$ model, we have obtained the exact solution of the field equations, presented in Eq.~\eqref{eq:exact_powerlaw_scale_factor}. We have also obtained the condition $\alpha+\beta>\frac{3}{2}$, which corresponds to an accelerating phase of the Universe. More generally, the exact solution allows us to recover distinct cosmological phases for different values of the combination of model parameters $\alpha+\beta$. In particular, $\alpha+\beta=\frac{1}{2}$, $\alpha+\beta=\frac{3}{4}$, and $\alpha+\beta=1$ correspond to the stiff-matter, radiation-dominated, and matter-dominated phases, respectively.

After determining the solution for the model, we have also validated it using a cosmological observational dataset. In this study, we considered various combinations of CC, PN$^{+}$, PN$^{+}$\& SH0ES, and the DESI DR2 dataset. For the different dataset combinations, we identified the posterior distribution along with the corresponding 68\% and 95\% confidence intervals. We calculated the optimal fit value for the model parameter alongside the Hubble constant $H_0$, $\Omega_{m0}$, and the nuisance parameter $M$. With the CC+PN$^{+}$ dataset, we derived $H_0 = 74.4^{+3.8}_{-4.1} \,\text{km s}^{-1} \, \text{Mpc}^{-1}$, which is a higher value compared to that of the $\Lambda$CDM for the same combination of datasets. Subsequently, after incorporating the SH0ES data points along with the CC+PN$^{+}$, we obtained $H_0 = 72.70^{+0.91}_{-1.09} \,\text{km s}^{-1} \, \text{Mpc}^{-1}$, closely resembling the standard $\Lambda$CDM model. These results are consistent with the $H_0$ value reported by the SH0ES distance-ladder measurement, which states $H_0 = 73.04 \pm 1.04 \,\text{km s}^{-1} \, \text{Mpc}^{-1}$ \cite{Riess_2022panplus}. For the combination of the CC+DESI DR2 dataset, we found $H_0 = 67.5^{+1.4}_{-1.6} \,\text{km s}^{-1} \, \text{Mpc}^{-1}$, which aligns with the Planck 2018 $\Lambda$CDM constraint of $H_0 = 67.4 \pm 0.5 \,\text{km s}^{-1} \, \text{Mpc}^{-1}$ \cite{Planck:2018vyg}. The inferred \(H_0\) is highly dependent on the dataset, ranging from early-Universe measurements to locally calibrated determinations.  

To evaluate how well the proposed model aligns with the standard $\Lambda$CDM model across various data set combinations, we compute the AIC and BIC. The associated $\Delta$AIC and $\Delta$BIC values for each data set combination are shown in Table~\ref{model_outputAICBIC}. Specifically, for the CC + DESI DR2 combination, both $\Delta$AIC and $\Delta$BIC are relatively low, suggesting that this combination offers a statistical fit similar to that of the traditional $\Lambda$CDM model when compared to other data set combinations. We have also analyzed the posterior covariance and correlation matrices to further explore parameter degeneracies and the reliability of the MCMC constraints. These matrices illustrate the strength and direction of correlations among model parameters, which align with the parameter degeneracies identified in the associated two-dimensional posterior contours. The covariance structure offers insights into the joint variations of the parameters, while the normalized correlation coefficients allow the relative strength of these degeneracies to be evaluated independently of the parameter scales.

To delve into late-time cosmology, we illustrate the evolution of fundamental cosmological parameters, such as the deceleration parameter and the total EoS parameter, for our selected \( f(Q, B) \) model in relation to the conventional \( \Lambda \)CDM model. We have also determined the current values of the deceleration and EoS parameters for each dataset combination. Our results are consistent with cosmological observations \cite{Gruber_2014, PhysRevD.90.044016a, PhysRevResearch.2.013028}. For each dataset combination, the deceleration parameter indicates a shift from a decelerating phase to an accelerating phase and, currently, reflects the Universe's accelerating expansion. The EoS parameter reveals varied cosmological regimes across different redshift intervals, highlighting distinct dynamical phases in the Universe's evolution. At higher redshifts, the reconstructed effective EoS becomes increasingly positive, indicating a stiff-like effective behavior of the background solution. However, this high-redshift behavior should be interpreted cautiously because the reconstruction is obtained from a numerical backward integration initialized at \(z=0\), while at low redshift dark energy takes over, reproducing the standard transition to accelerated expansion. We have also compared the present day value of the Hubble constant values derived from various findings in the literature. Our results are consistent with the existing studies.  
%%%%%%%%%%%%%%%%%%
\section*{Appendix}

\begin{figure}[htb]
 \centering
 \includegraphics[width=85mm]{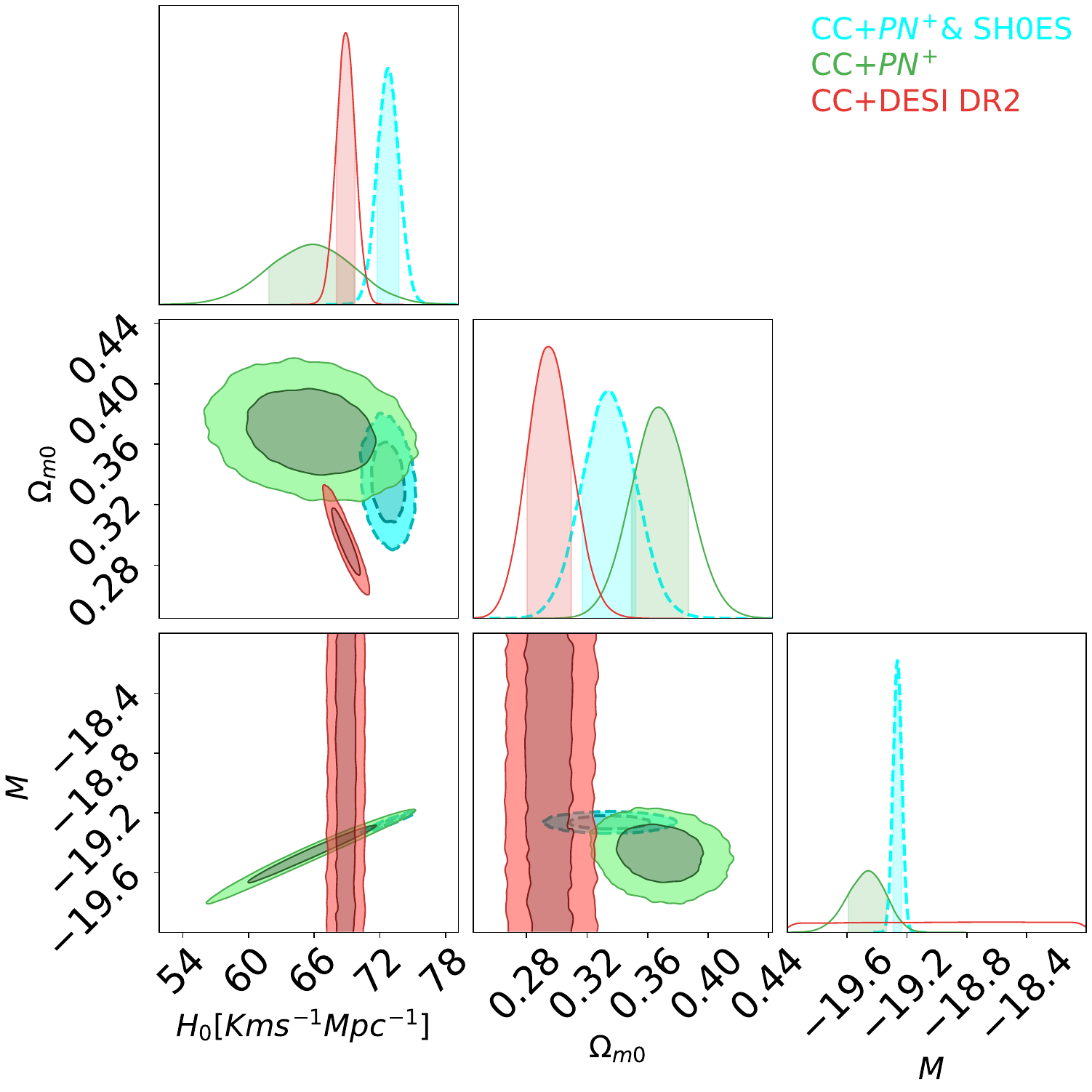}
 \caption{The marginalized posterior distributions and the corresponding 68\% and 95\% confidence contours for the $\Lambda$CDM model parameters, derived from the different dataset combinations.} \label{plot:FigCDMMCMC}
 \end{figure}

We present findings based on the $\Lambda$CDM model, emphasizing the posterior distributions depicted in Fig. \ref{plot:FigCDMMCMC}. This figure shows the $1\sigma$ and $2\sigma$ confidence intervals for different dataset combinations, providing a comprehensive analysis of parameter constraints. A summary of the detailed results for each dataset combination is provided in Table \ref{LCDM_outputs}. This approach highlights the alignment of our results with the standard predictions of the $\Lambda$CDM framework, thus enabling a stringent evaluation of the model's efficacy across the chosen datasets.

\begin{table}
\renewcommand{\arraystretch}{1.5}
    \centering
    \caption{The table presents results for the $\Lambda$CDM  model, with the first column listing the data set combinations. The second column shows the constraints on the Hubble constant $H_0$, while the third and fourth columns present the values for the matter density $\Omega_{m0}$ and nuisance parameter $M$ respectively. The fifth, sixth and seventh columns display the statistical terms $\chi^2_{min.}$, AIC and BIC respectively.}
    \label{LCDM_outputs}
    \begin{tabular}{ccccccc}
        \hline
		Data sets & $H_0\,\text{km} \, \text{s}^{-1} \text{Mpc}^{-1}$ & $\Omega_{m0}$ & $M$&$\chi^2_{min}$ & AIC & BIC\\ 
		\hline
        CC+$PN^{+}$ & $66.0^{+3.7}_{-4.1}$ & $0.367^{+0.020}_{-0.017}$ & $-19.46^{+0.12}_{-0.14}$ &1792.30 &  1798.30& 1814.67\\
		CC+$PN^{+}\&$ SH0ES & $72.76^{+0.96}_{-1.03}$ & $0.335^{+0.017}_{-0.018}$ & $-19.264^{+0.031}_{-0.028}$ &1539.22  & 1545.22 & 1561.59\\ 
		CC + DESI DR2 & $68.87^{+0.87}_{-0.83}$ & $0.294^{+0.016}_{-0.013}$ & Unconstrained& 19.08& 25.08 & 30.36\\ 
		\hline
    \end{tabular}
\end{table}

 % Adjust this value for more or less space

\section*{Acknowledgments} 
Y.S. and L.K.D. acknowledge the support of the National Natural Science Foundation of China under  Grant Nos. T2241005  and 12075059, as well as the startup fund of USTC.

%\section*{References}

\bibliographystyle{spphys}   
\bibliography{bio}   

\end{document}